\documentclass[journal,onecolumn]{IEEEtran}
\usepackage{graphicx}
\usepackage{amsmath}
\usepackage{enumerate}
\usepackage{multirow}
\usepackage{epstopdf}
\usepackage{array}
\usepackage{CJK}
\usepackage{float}
\usepackage{subfigure}
\usepackage{wrapfig}
\usepackage{algorithm}
\usepackage{algorithmicx}

\usepackage{algpseudocode}
\usepackage[justification=centering]{caption}
\renewcommand\thesection{\arabic{section}} \renewcommand\thesubsection{\thesection.\arabic{subsection}}

\begin{document}

\title{Data-driven Computational Social Science: \\A Survey}
\author{Jun Zhang, Wei Wang, Feng Xia, Yu-Ru Lin, Hanghang Tong
\thanks{J. Zhang is with Graduate School of Education, Dalian University of Technology, Dalian 116024, China}
\thanks{W. Wang is with Department of Computer and Information Science, University of Macau, Macau 999078 China}
\thanks{F. Xia is with School of Engineering, IT and Physical Sciences, Federation University Australia, Ballarat, VIC 3353, Australia}
\thanks{Y.-R. Lin is with School of Information Sciences, University of Pittsburgh, Pittsburgh, PA 15260, USA}
\thanks{H. Tong is with Department of Computer Science, University of Illinois at Urbana-Champaign, Urbana, IL 61801, USA}
\thanks{Corresponding author: Feng Xia; Email: f.xia@ieee.org}
       }
\maketitle

\begin{abstract}
Social science concerns issues on individuals, relationships, and the whole society. The complexity of research topics in social science makes it the amalgamation of multiple disciplines, such as economics, political science, and sociology, etc. For centuries, scientists have conducted many studies to understand the mechanisms of the society. However, due to the limitations of traditional research methods, there exist many critical social issues to be explored. To solve those issues, computational social science emerges due to the rapid advancements of computation technologies and the profound studies on social science. With the aids of the advanced research techniques, various kinds of data from diverse areas can be acquired nowadays, and they can help us look into social problems with a new eye. As a result, utilizing various data to reveal issues derived from computational social science area has attracted more and more attentions. In this paper, to the best of our knowledge, we present a survey on data-driven computational social science for the first time which primarily focuses on reviewing application domains involving human dynamics. The state-of-the-art research on human dynamics is reviewed from three aspects: individuals, relationships, and collectives. Specifically, the research methodologies used to address research challenges in aforementioned application domains are summarized. In addition, some important open challenges with respect to both emerging research topics and research methods are discussed.
\end{abstract}


\IEEEpeerreviewmaketitle

\section{Introduction}

\IEEEPARstart{H}{uman} dynamics is proliferated with social interactions and permeates all facets of human existence, including education, commerce, health consultations as well as expressing emotions to intimate peers and so on. These multifarious encounters often leave residual traces that may either be online or offline. Online data can be mined and extrapolated through existing scientific and mathematical techniques. These digital footprints often can accurately reflect the mechanisms of the real world. However, due to technological limitations and privacy issues, previous research can not collect and process related data with ease. Thanks to the efforts of scientists and the advanced processing technologies, detailed personal data are now available such as relationships, GPS coordinates, community memberships, and contact frequency. Exploiting such data can provide a new perspective for garnering invaluable insights into human social phenomena.

Social science has a long history which includes many areas such as anthropology, economics, political science, psychology, and sociology etc. It investigates issues in the above disciplines or inter-disciplines by using theoretical analysis and experimental results. The scale of problems that social scientists study ranges from micro to macro level. For instance, economists can explore individuals' investment behaviors or predict global economic changes by using financial datasets or data collected by mobile phones \cite{toole2015tracking}.

Traditional methods of retrieving empirical data for the analysis of issues in the domain of social science are founded on the principles of social investigations. These procedures are often conducted with manual retrieval using questionnaires through human resources. This is not only quite resource-intensive, but also leads to gross inaccuracies either due to human negligence or other inherent limitations. Therefore, Computational Social Science (CSS) emerges as time requires, which takes advantage of mathematical theories together with data processing and analyzing technologies  \cite{liu2018accessai, liu2018accessvis} from computational science to tackle those social issues.

Nowadays, together with the help of the above mentioned advanced research techniques, the accessibility of the diverse data on human beings significantly influences the research topics or methods that researchers concern. Through utilizing such data, traditional social issues can be investigated from a new perspective. More social phenomena can be discovered. Meanwhile, new research topics or methods emerge because of the availability of data. Therefore, exploring the data-centered research topics in computational social science area has attracted more and more attentions. Human is the most fundamental component of our society, and the profound understanding of human dynamics has always been the goal of scientists with diverse backgrounds. With the emergence of CSS, we now can look into the research issues on human dynamics from a data-driven aspect. Human dynamics is the main focus of this paper. Human dynamics refer to a branch of complex systems research in statistical physics such as the movement of crowds and queues and other systems of complex human interactions including statistical modelling of human networks \cite{Hayat2020csr, xia2019accessmotif}, including interactions over communications networks. Inspired by the above observations, we present a survey on data-driven computational social science. The overall research topics on human dynamics are reviewed from the following three perspectives: individuals, their interacting relationships, and collectives.

Individualistic analysis mainly focuses on individual attributes (e.g., basic human features and personal influence) and individual behaviors (e.g., basic human actions and behavior prediction). Individuals can be organized into communities based on similar attributes or similar relationships. Each individual in the communities can have an impact on the collectives. On the opposite, communities can influence the individuals inside them. These mutual effects attract many scientists to explore the mechanisms between individuals and communities. As a result, research on individuals, their links to others, and collectives based on real-world's data are of great significance to our whole society.

\begin{wrapfigure}{R}{0.5\textwidth}
  \begin{center}
    \includegraphics[width=3.6in]{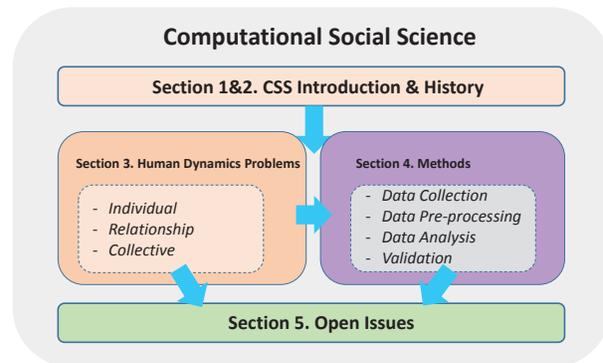}
  \end{center}
  \caption{Architecture of the survey.}
  \label{fig-1}
\end{wrapfigure}

Research on relationships involves a discussion on the identification of different types of relationships between individuals. Based on individuals' predispositions and their historical interactions, predictions of newly appeared relationships and vanished interactions can be made. The relationships between individuals are constantly evolving due to the ever-changing environment as time goes on. Special consideration is given to relationship dynamics employing big data which offers a moment-by-moment picture of interactions over extended periods of time. Finally, the mechanics of temporal evolution are expounded upon using periodic snapshots in human communications.

Collective analysis focuses on community detection, evolution, and collective behaviors. Several typical community detection methods are presented, including modularity-based methods, social interaction based methods, and overlapping community detection algorithms. Key issues in community evolution are summarized and presented. Additionally, several emerging trends in collective behaviors are discussed, including computational social choice, cooperation, human mobility, social contagion, and collective behavior prediction.

Besides reviewing emerging research topics on human dynamics, we also summarize the procedure of how to solve those issues. Data collection and pre-processing are the initial steps in social science research. Then, we mainly focus on the procedure of data analysis with the combination of traditional statistics methods and the most widely employed machine learning methods. At the final stage, the validation process is needed in order to ensure correctness and accuracy. According to the above-mentioned steps of our research methodology, which are data collection, data pre-processing, data analysis, and data validation, we then expound the most frequently used approaches in each stage in detail.

It has been a decade since the pioneering research on computation social science \cite{lazer2009life}. Significant research progresses on computational social science have been made by scholars from diverse areas. In this paper, we review the evolvement of research topics, methodologies, and challenges as comprehensive as possible in computational social science area. Generally speaking, we make the following contributions in this paper:

\begin{itemize}
\item To the best of our knowledge, this study is the first to survey the application domains involving human dynamics in the data-driven computational social science area.
\item The state-of-the-art research topics are elaborately introduced from three perspectives: individuals, relationships, and collectives.
\item Specifically, research procedures and methodologies that can be utilized to solve the above mentioned human dynamics issues are summarized.
\item The future trends and open challenges are discussed with respect to both research topics and methodologies in this field.
\end{itemize}

The organization of this paper is shown in Fig.~\ref{fig-1}. The history of CSS is immediately forthcoming in Section 2. Section 3 presents the research milestones from the perspectives of individuals, relationships, and collectives. The methods of processing data are discussed in Section 4. Then, open issues are put forward and discussed in Section 5. Finally, the paper is concluded in Section 6.

\section{Computational Social Science and Related Areas}

Social science has a long history which aims to investigate human society at all levels in terms of individuals, communities, organizations, societies, and the whole world. Social scientists have developed a vast body of knowledge for understanding human behaviors and social systems since the 18th century. Traditional social science mainly consists of social psychology, anthropology, economics, political science, and sociology, which are the so-called ``Big Five". Each of these five disciplines has many other specific branches. The traditional social science in this formative age is adopted as a scientific culture concerning the quest for and understanding human behaviors.

\begin{figure*}
\centering
\includegraphics[width=5.5in]{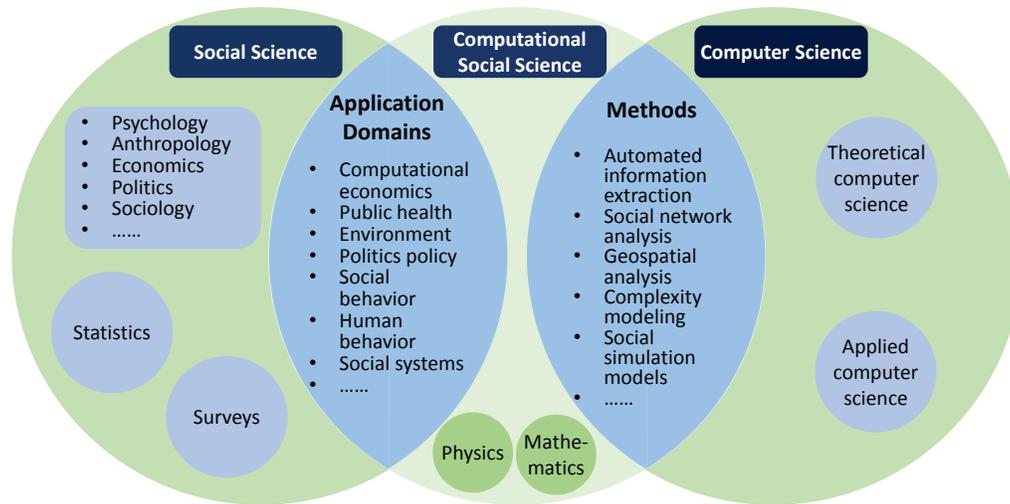}
\caption{Interdisciplinary nature of CSS.}
\label{fig.2}
\end{figure*}

The traditional methods of social science mostly are scratching the surface of human society, while new tools can shed light on social issues from totally different angles. The origination of CSS can be dated to the 1960s, when social scientists began using computers for analyzing statistical data. According to Cioffi-Revilla~\cite{cioffi2010computational}, the beginnings of CSS can be marked, in a strict sense, with the invention of digital computing from the early days of the Cold War. The modern digital computation methods are the basis of the new CSS area by providing the key instrument that would vastly accelerate computing in a way that would have seemed incredible just a few years ago. Accordingly, CSS is by all means an instrument-enabled scientific discipline. Just like Galileo Galilei invented the telescope as the key instrument for observing and understanding the physical universe, computational social scientists try to exploit the advanced and powerful computation instrument to reanalyze the traditional social science discipline. For the first time, social scientists (or computational social scientists) are able to analyze the large volume of data in order to test previous scientific social hypotheses. For example, the early CSS pioneers such as Charles Spearman, Rudolf Rummel, and L. Thurnstone proposed the many powerful theories that led to many remarkable discoveries across social science~\cite{cioffi2014introduction}.

Recently, the capacity to collect and analyze massive amounts of data has promoted the evolution of CSS and expedites the emergence of data-driven CSS. Just as Lazer et al. said in \cite{lazer2009life}, we now live life with many digital traces. We communicate with our friends through online social media, make mobile phone calls from anywhere, use public transportation with a transit card, and make purchases with a digital account. We may go out by taxi, from which our moving traces are recorded by GPS devices. CSS could take full advantage of all these digital traces to better understand the individual or collective behaviors and furthermore to better understand our society and solve social problems.

An important feature of CSS is the interdisciplinary nature. CSS is an integrated and interdisciplinary new area, which aims to analyze previous and present social issues and human behaviors with an emphasis on information processing through the methods of advanced computation. As we can see from Fig.~\ref{fig.2}, the two main disciplines composing CSS are computer science and social science. CSS mainly focuses on analyzing the issues studied by social scientists with the computational methods which have been developed by computational scientists. The field is inherently interdisciplinary. Social scientists provide the insight into pertinent research issues such as economics, politics, and environment, while the computer scientists contribute expertise in developing mathematical methods such as social network analysis, data mining, and machine learning. A useful clarification to keep in mind is that CSS is not limited to Big Data, or Social Computing, or any other relatively areas. In short, CSS is a fundamental interdiscipline including all social science disciplines, applied computer science, and other related disciplines.

CSS has huge potentials for applications in understanding human behaviors and social dynamics from various aspects. Lazer et al. \cite{lazer2009life} have discussed the issues that CSS intends to answer from human interactions to social networks. For example, group interactions could be examined through online social media data. CSS can also address questions about the temporal dynamics of human behavior (e.g. what is the individual mobility pattern? how does human relationship evolve? how to detect human community from social media data?) \cite{guo2016mobigroup, ozdikis2016evidential}. Scholars in this emerging area can use related data to solve these social problems with advanced computing technologies. Meanwhile, CSS has many application fields such as economics, sociology, geography, public health, environment, and political science, as shown in Table~\ref{tab:CSS}.

\begin{table*}[htbp]
\centering
\caption{Application areas of CSS.}
\label{tab:CSS}
\begin{tabular}{|m{3cm}<{\centering}|m{5cm}<{\centering}|m{5cm}<{\centering}|}                                               \hline
Application areas   & Goals   & Methods               \\ \hline
Economics        & Understand people behavior in economic activities; personal recommendation service       &Agent-based model; recommendation technology                 \\ \hline
Sociology        & Analyzing human interaction; social behavior and social systems        &Social simulation modeling; social network analysis                 \\   \hline
Geography        & Human mobility analysis; urban sensing; location-based recommendation service        &Complexity modeling; automated information extraction                     \\ \hline
Public health    & Anticipate and track disease outbreak; understand the spread of disease and prevention       &Statistic; social simulation modeling    \\ \hline
Environment      & Nature preservation; pollution measurement         &Data mining; big data analysis                           \\ \hline
Politics         & Understand human voting behavior; party opinion diffusion analysis          &Computational social choice   \\ \hline
\end{tabular}
\end{table*}

\section{Research Topics}

\begin{wrapfigure}{R}{0.5\textwidth}
  \begin{center}
    \includegraphics[width=3.6in]{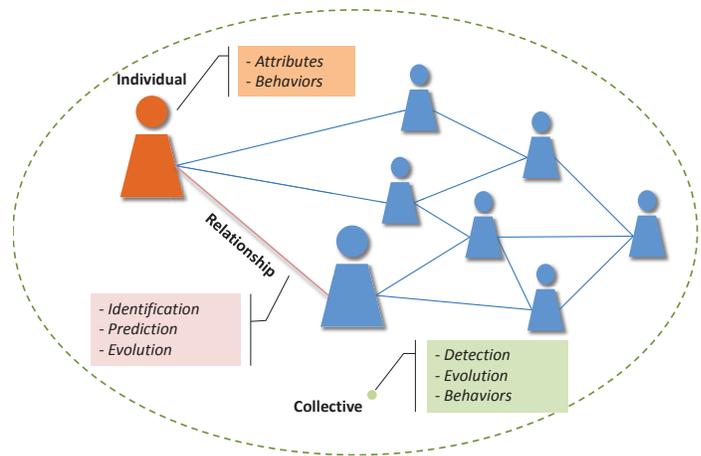}
  \end{center}
  \caption{Involved topics of human dynamics.}
  \label{fig3}
\end{wrapfigure}

Sociologists and engineers discover a majority of laws or techniques to reveal the nature of our society from the huge amount of data obtained via various resources. Such findings can help to accelerate the advancements of our society. The fundamental components of human society are human beings, the links among them, and the ``contagion" carried on those links \cite{christakis2009connected}. Discovering the mechanisms and attributes of those basic components and taking full advantage of them can greatly benefit human beings in understanding the whole society with a fairly comprehensive view. Therefore, we unfold the above basic segments from three general views which are individuals, relationships, and collectives. The main components of the above three parts are described in Fig.~\ref{fig3} and Table ~\ref{tab:all}.

\begin{table*}
\centering
\caption{Research topics in human dynamics area.}
\label{tab:all}
\begin{tabular}{|m{3cm}<{\centering}|m{5cm}<{\centering}|m{5cm}<{\centering}|}                                               \hline
Research topics& Sub-categories & Specific Topics and Citations      \\ \hline
Individual & Individual Attributes and Behaviours     &Personal Features [8-12] ; Individual Influence [13-18];   Human Actions [19-24]; Influence Factors [25-30]; Behavior Prediction [31-33]              \\ \hline
Relationship & Temporal Evolution of Relationship         & Relationship Identification [34-42]; Relationship Prediction [43-61];Relationship Evolution [62-64]              \\   \hline
Collective & Community and Behaviours      &Community Detection [71-83]; Community Evolution [84-87]; Community Behaviours [89-147]                   \\ \hline
\end{tabular}
\end{table*}

\subsection{\textbf{Individuals}}
\begin{wrapfigure}{R}{0.5\textwidth}
  \begin{center}
    \includegraphics[width=2.7in]{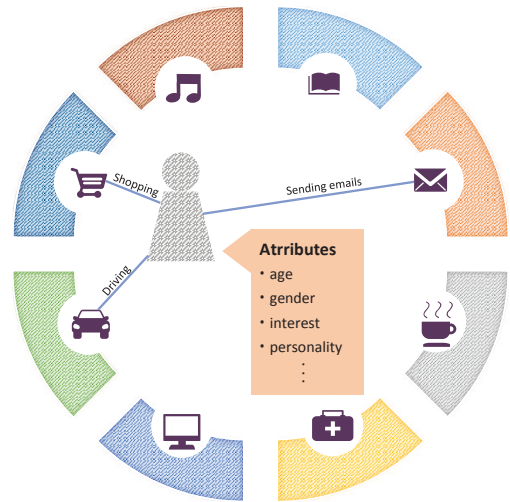}
  \end{center}
  \caption{An illustration of individual analysis.}
  \label{individual}
\end{wrapfigure}

As previously stated, a growing proportion of human activities can be easily accessed from online social networks and mobile sensor devices, which provide multifarious data sets and make rapid progress on the research in CSS. To understand how the developments of CSS contribute to the research of human behaviors and social dynamics, we take individual analysis as the point of penetration, since individuals are the basic elements of the whole social network. As shown in Fig.~\ref{individual}, people have their personal attributes, such as age, gender, interest, and personality, etc. Meanwhile, they also have a series of activities, e.g., driving, shopping, sending emails to others, etc. In order to make a better understanding of human beings from the individual level, we must take both individual attributes and individual behaviors into account.

%\subsubsection{Individual Attributes}

\textit{1) Individual Attributes}

Every person has particular attributes. On one hand, some of these attributes are intrinsic and have no relations with other people, such as gender, religion, and personality. These personal features can be found through mining the relative data. On the other hand, people have an influence on each other as a part of society. Influence, which measures the individuals' effects on others and takes consideration of the whole network, is an important social attribute of individuals.

\textit{a. Personal Features}

Our daily lives are recorded by digital services. For example, social media records what we are saying and mobile sensing devices record where we are. These digital traces consist of information on a large range of subjects which often correlate to personal properties of individuals, including physiological features and personality. Besides, other attributes such as human interests and human locations, which also reflect individuals' personal information, are utilized for research purposes.

Information gathered from online social media can be used to analyze the physiological features of individuals, including gender, age, race, etc. Park et al. \cite{park2016women} analyzed 10 million messages from over 52,000 Facebook users to explore the differences in language use across gender, affiliation, and assertiveness. Some other highly subtle personal attributes such as ethnicity, religious, political views, even sexual orientation, also can be accurately predicted using social media records.

Personality can be reliably evaluated by combining the digital trail with the Five-Factor Model \cite{digman1990personality}, which has a distinguished reputation for the most accurate prediction of personality traits. In this respect, de Montjoye et al. \cite{de2013predicting} were among the first to show evidence that personalities could be reliably predicted using mobile phone carrier logs. In the study of individuals' interest information, Zhao et al. \cite{zhao2013emergence} investigated the dynamics of human interests such as the duration of continuous interest, the return time of visiting a certain interest, interest ranking, and interest transition by a biased random walk model. This was done using two e-commerce datasets and one communication dataset. For the human location research, Zheng et al. \cite{zheng2017survey} offered an overall picture of location prediction on Twitter. The Twitter network, tweet content, and tweet context are used as potential inputs to predict user home locations, tweet locations, and mentioned locations, then analyze how these problems depend on these inputs.

\textit{b. Individual Influence}

Influence is an important measurement to indicate individuals' positions in the community. Identifying the most influential individuals (nodes) is critical in understanding the dynamics of a social network. Quantifying the influence of an individual on others has become more accurate as massive amounts of social context are increasingly available \cite{ren2019accessapi}.

The main method to quantify individuals' influence is through network analysis. Centrality \cite{bonacich1987power, zhang2019wwwcover}, which is a traditional network analysis method, is a measure to assess the importance of nodes in the network. Now researchers are putting efforts to develop new centrality computing methods. Sun et al. \cite{sun2017identifying} proposed an efficient algorithm for identifying influential nodes, using weighted formal concept analysis and simulated on several real-world networks.

There also exist other methods targeting at finding the most influential people in an online social network other than the centrality method \cite{neshati2017dynamicity}. For instance, Weng et al. \cite{weng2010twitterrank} considered the homophily of users with a similar focus. They extended a modified PageRank algorithm, entitled TwitterRank to measure the influence of users on Twitter. Subbian et al. \cite{subbian2016mining} proposed a novel algorithm InFlowMine to discover the information flow patterns in the network to find the top influencers, and demonstrate the effectiveness of the discovered information flows using an influence mining application. Additionally, considering the mechanism of the Weibo social network, Hong et al. \cite{hong2017user} developed quantitative measurements for user vitality ranking algorithm, which considers the mutual influence between users. Other than user vitality ranking, they also predict the importance of users.

\textit{2) Individual Behaviors}

Attributes analysis is not enough to describe individuals' characteristics. Whether as an independent individual or as a part of society, people need various activities. Therefore, behavior analysis is another topic related to research on human beings. In this section, we introduce the behavior analysis at the individual level, such as what kind of actions researchers are focused on, what factors can influence individuals' behaviors, and how to predict individuals' behaviors.

\textit{a. Human Actions}

Our basic actions, like sentiment, mobility, even routine behaviors, leave traces on websites and electronics devices. This phenomenon has led to a boom in the investigation of human activity discovery and human action comprehension.

Sentiment analysis aims to extract individuals' subjective information, such as evaluation of something, emotion states, and polarity of attitude. Nowadays, sentiment analysis has been quite popular as it has a variety of applications in the fields of marketing, politics, online shopping, amongst several others. Ortigosa et al. \cite{ortigosa2014sentiment} developed a new method combining lexical-based and machine-learning techniques for the extraction of user positive/neutral/negative sentiments and the detection of user sentiment changes. Hu et al. \cite{hu2013unsupervised} investigated a novel problem of polarity sentiment classification with emotional signals including emotions, product ratings, restaurant stars, etc., and evaluated the method on two Twitter datasets, i.e., Stanford Twitter Sentiment and Obama-McCain Debate.

Most human activities are connected with movements. Individuals' mobility pattern is an important aspect for understanding human behaviors \cite{xia2020tits, xia2018tiiurban}. Gonzalez et al. \cite{gonzalez2008understanding} and Song et al. \cite{song2010modelling} investigated the properties of human mobility by studying the users' trajectories captured by mobile phone traces. Emphatic research has been conducted on the prediction of human mobility, which will be expounded later.

Routine behaviors can also be investigated to some extent, which may provide further insight into human behaviors. Barab$\acute{a}$si \cite{barabasi2005origin} showed that the timing of many human actions follows heavy-tailed distribution based on an e-mail data set which captures the information of the sender, recipient, time, and the size of each e-mail. Farrahi et al. \cite{farrahi2010probabilistic} proposed a probabilistic methodology to mine meaningful details about human routine activities, for example, being at home in the morning and being out on Friday evening, through integrating human interaction data (obtained by mobile phone bluetooth sensors) with human location data (obtained by mobile cell tower connections).

\textit{b. Influence Factors}

After analyzing basic human actions, we need to comprehend what factors can influence these actions and how these different factors affect individual behaviors. Researchers aim to discover these determining factors, such as individual characteristics, sites, or environmental aspects, which are related to individual behaviors and behavior intensity.

The differences in various website platforms are such an influential factor. Tsou et al. \cite{tsou2014community} analyzed the commenting behavior of users for two video sharing websites, i.e., YouTube and TED (Technology, Entertainment, Design). The results indicated significant differences in the observed commenting behaviors according to platforms and presenters' characteristics. Smith et al. \cite{smith2012does} compared user-generated content (UGC) across three social media sites (Twitter, Facebook, and YouTube) to address the question of how users engage with different social media platforms and how brand-related UGC varies across different social media types.

Some other research also assists in describing the factors that influence individual behaviors. Focusing on social context influence and using a database obtained from the ``Diet and Fitness" category of Yahoo Answers, Kuebler et al. \cite{kuebler2013overweight} explored an examination of obesity which suggested that obese people who lived in counties with higher than average Body Mass Index (BMI) had greater safeguards from mental and physical health ailments. Escobedo et al. \cite{escobedo2017shine} proposed the SHINE-L system which uses acoustic sensors of mobile phones/smartwatches to detect and record family routines, including family meal frequency, sleep, and screen time for social learning.

Besides investigating factors that have impacts on individual behaviors, scientists also explore the external causes of human's various sentiments. Based on a dataset from millions of public Twitter messages, Golder et al. \cite{golder2011diurnal} identified how diurnal and seasonal mood varied with work, sleep, and daylength at an individual level. Yang et al. \cite{yang2016life} examined trends in posts about life satisfaction from a two-year sample of Twitter data. They applied a surveillance methodology to extract expressions of both satisfaction and dissatisfaction with life.

\textit{c. Behavior Prediction}

The accessibility of huge datasets and the advanced CSS techniques allow for greater flexibility and accuracy when predicting human behaviors. One of the phenomena that reflects on the individual level is the mobility and location prediction of individuals.

Entropy is a fundamental quantity capturing the degree of predictability. By this measurement, Song et al. \cite{song2010limits} explored the limits of predictability in human mobility and found an exceptionally high value of potential predictability using a trajectory data set captured from mobile phones. Considering the correlations between individuals, Zhang et al. \cite{zhang2016gmove} proposed a group-level mobility modeling method that utilizes the geo-tagged social media data. All the above studies were devoted to research individuals' offline mobility. Online mobility can also be well predicted as well as offline mobility. Using two datasets from the online virtual universe Pardus (www.pardus.at), which contained users' behavioral actions and trajectories, Sinatra et al. \cite{sinatra2014entropy} proved that behavioral actions in virtual human lives showed high levels of regularity and predictability with three entropy computing methods.

\subsection{\textbf{Relationships}}
\begin{wrapfigure}{R}{0.5\textwidth}
  \begin{center}
    \includegraphics[width=3.6in]{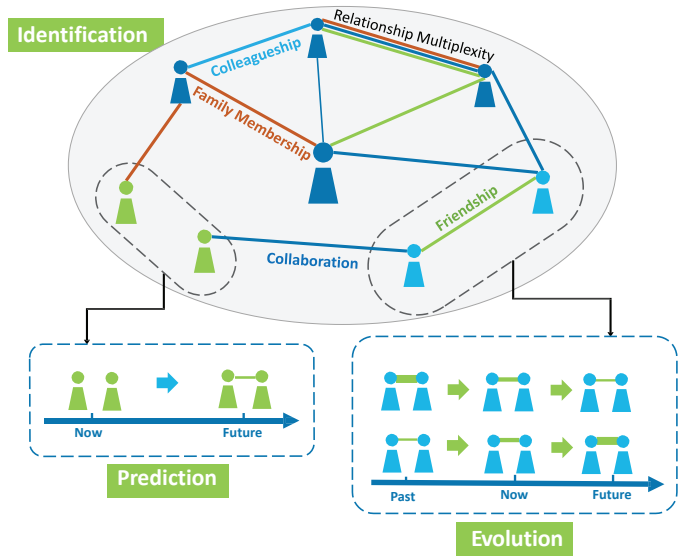}
  \end{center}
  \caption{An illustration of relationship dynamics.}
\label{relationship dynamics}
\end{wrapfigure}

Individuals are embedded in social networks and are therefore intrinsically connected to a neighboring group of people \cite{xu2020jcdl, liu2020sigir}. These relationships can be formed through online communication or offline interactions. Relationships between individuals are constantly evolving due to continuous changes in interactions and behavior. By analyzing and partitioning socially relevant big data, researchers can glean a moment-by-moment representation of both, the structure and content of relationships. To this end, distinct types of relationships between individuals can be identified through their interactions and social network structure. Moreover, based on historical network information and individual features, future relationships can be predicted. Recently, significant interest is being placed on the mechanisms used to capture the temporal evolution of relationships using these periodic snapshots of human communication. These processes of relationship dynamics are shown in Fig.~\ref{relationship dynamics}, including relationship identification, prediction, and evolution.

\textit{1) Relationship Identification}

In the ambit social network, each individual possesses a network neighborhood, which can be defined as the set of people with whom he/she has some forms of relationship. It shows the variety of social ties which encompass a profoundly diverse set of relationships. Such relationships typically include family members, classmates, colleagues, friends of long duration, distant acquaintances, co-authors, and a variety of other categories. A crucial and broad issue for the analysis of social networks is to leverage features in the available data to mine the relationship semantics and identify the type of relationships.

There are many papers that focus on the identification of intimate relationships such as friendship and romantic involvement. Steurer et al. \cite{steurer2013acquaintance} conducted experiments using several methods, including a statistical approach, supervised and unsupervised learning algorithms, and leveraging online social network data and location-based data to classify the relationship as either partners or acquaintances. Backstrom et al. \cite{backstrom2014romantic} defined a novel measure of tie strength as ``dispersion" to recognize one's spouses or romantic partners. This method used data from a large sample of Facebook users, and the term dispersion represents the extent to which two people's mutual friends are not themselves well-connected. Other researchers used sensor technology to design experiments and collect offline interaction data. Oloritun et al. \cite{oloritun2013identifying} employed logistic regression on time-resolved face-to-face interactions, to identify close friendship ties via interactions at different periods, spatial proximity, and gender similarity. It was found that residence proximity and duration of weekend night interactions may help to explain close friendship ties. Previous work has mainly focused on binary friendship relations, which provides only a coarse representation of relationship identification. However, the low cost of relationship formation in social networks can lead to a distribution of heterogeneous relationship strengths. Ju et al. \cite{ju2017novel} developed an online social networks users relationship strength estimation model that incorporates topic classification and indirect relationship through the evaluation of Weibo datasets.

Some researchers try to identify relationships in an organization, for example, companies and laboratories. Tang et al. \cite{tang2012inferring} incorporated social theories into a factor graph model to classify the type of social relationships by learning across heterogeneous networks, and assessed its performance on a manager-subordinate relationship identification task using the email corpus of Enron. Choi et al. \cite{choi2013mining} found indicators from communication patterns including co-location data and instant messenger data to infer social relationship types, which were divided into formal and informal. In addition, Wang et al. \cite{wang2017shifu} proposed Shifu, a deep-learning-based advisor-advisee relationship identification method that takes into account both the local properties and network characteristics. Liu et al. \cite{liu2019tkdeshifu2} devised a novel model based on network representation learning, namely Shifu2, for advisor-advisee relationship mining. More generally, Rotabi et al. \cite{rotabi2017detecting} used structural network features for detecting strong ties with high precision via experiments on Twitter data.

Individuals engage with many overlapping social relationships and have diverse social roles across different facets of their lives. Moreover, the daily life of a single individual always assumes multiple roles. Therefore, individual relationships in our daily life are comprised of multiple facets, for example, a family member also belongs to a circle of friends. This dichotomy is termed as relationship multiplexity. According to life facet (\emph{family, work,} and \emph{social}), Min et al. \cite{min2013mining} introduced machine learning techniques to classify the distinguished contacts, using profile information as well as five categories of communication behaviors, including intensity, regularity, temporal tendency, channel selection/avoidance, and maintenance cost. Mining the semantics of relationship multiplexity makes social network analysis plentiful and closer to our real physical social networks. In addition, it would be also useful to take advantage of the semantic information regarding inferred relationships to help us manage our social lives.

\textit{2) Relationship Prediction}

Due to the unprecedented interest in relationship dynamics, relationship prediction has attracted significant attention. This problem is approached by predicting the emergence of relationships based on predefined features and interactions, as well as to infer when relationships will be established. The relationships between individuals are diverse and overlapping, but friendship has been found as a key factor that maintains users active in social networks and the whole community expanding. Hence, the majority of researchers are dived into friendship prediction employing machine learning and statistical methodologies. At the same time, other relationships have also been studied, such as co-authorship and trust relationships. Undoubtedly good prediction mechanisms can facilitate friendship recommendation, the precaution of disease propagation and advertisements, etc.

In friendship prediction, homophily is an emerging research hotspot, which is traditionally accounted for people who are always bond with people alike. It suggests that the more similar users are, the greater the possibility they will be friends in the future. The formation of online friendships may reflect real-world encounters to some degree. In other words, people are more likely to become friends when they have mutual interests or are geographically close. For the similarity of interest, different research methods may at times lead to the paradoxical conclusion. Yang et al. \cite{yang2011like} proposed a joint friend-interest propagation framework by devising a factor-based random walk model, which achieved a greater performance in both interest targeting and friendship prediction. Parimi et al. \cite{parimi2011predicting} showed that the Latent Dirichlet Allocation (LDA) based interest features helped to improve the prediction performance when used in combination with graph features. To evaluate the power of friendship propagation in link prediction, Zhang et al. \cite{zhang2013learning} designed LFPN-RW which modeled friend-making behavior as a random walk upon the Latent Friendship Propagation Network. This naturally provided greater prediction accuracy, by capturing the co-influence effect of friend circles as well as personal interests.

On the other hand, location proximity and simultaneous occurrence at one place may not possess a good predictive power for the majority of friendship prediction. If the pair users are co-attendance at the same site frequently, they will be friends with high possibility. People are increasingly sharing their locations with their friends through the online location-based social network (LBSN). Such ``check-ins" can be broadcasted to friends, while information related to a place can be generated and shared. Therefore, location features are extracted from LBSNs for friendship prediction. Such features may include a combination of topological and geographical features, i.e., locations of individuals and their friends \cite{sadilek2012finding}, users' location trails over time \cite{cranshaw2010bridging}, as well as proximity and co-occurrence of periodic snapshots with temporal evolution \cite{crandall2010inferring}. These studies show a positive relationship between location information and social tie formation, which helps to improve the accuracy of inferring friendships.

The social proximity between individuals is driven not only by homophilous relations, but also by their interactions. Friendship prediction is strongly related to the interactions between users across the various forms of communication \cite{Hou2019prediction}. There is a higher possibility of becoming friends and intimate partners with greater the frequency of interactions. As a result, quite a high amount of research activities are devoted to discovering and predicting interactions between users. For instance, Steurer et al. \cite{steurer2013predicting} studied the extent to which interactions between users in online social networks can be predicted exploring features obtained from social network and position data from Second life social network. In addition, Nasim et al. \cite{nasim2016investigating} study the impact of additional interaction information on the inference of links between nodes in partially covert networks through data collected from 586 Facebook profiles.

At the same time, a great quantity of research focuses on friendship prediction by leveraging information on content and patterns of human interaction. The features extracted from individual interactions are from different knowledge sources, which include real-world sensing data from mobile phones, online social network exchange information, and the combination of online communication and offline interactions. Consequently, numerous researchers \cite{yu2013understanding} used features extracted from mobile phones, which included calls, text messages, and geographic propinquity, to infer friendships using supervised and unsupervised machine learning methods. Online exchange information was also employed to forecast friendships between users, containing communication attributes like link-sign, exchange messages indicating sentiments, and transactional information in the context of user behavior \cite{chen2013friendship, tang2016recommendations}. In addition, for the combination and comparison of online and offline links, Bischoff et al. \cite{bischoff2012we} contrasted online friendship with offline interactions through the physically co-attended events on Last.fm. It was found that self-similarity existed along with social ties and strong overlapping occurred in social circles increasing with tie strength. Castilho et al. \cite{castilho2014working} also collected data from offline team formation and online interactions, and found Facebook-derived proxies for tie strength, popularity, and homophily are more predictive of whom members wish to work within a class, compared to class grades.

The prediction of other types of relationships (such as trust and co-authorship relations) can be exploited in relationship dynamics. In \cite{he2015trust,tang2013exploiting}, researchers leveraged cluster-based classification method, using an unsupervised framework incorporating low-rank matrix factorization and homophily regularization respectively to predict trust relationships. These approaches were evaluated on real-world datasets from product review site -- Epinions. It was found that the number of active users and homophily helped to achieve greater accuracy. Along this line, co-author relationship prediction has been investigated. Mostly, authors are considered to belong to a homogeneous network, which means that only one type of object (authors) and one type of link (co-authorship) exist in the network. Yu et al. \cite{yu2014predicting} proposed a supervised model to learn the best weights associated with different topological features in deciding co-author relationships in the research field of Coronary Artery Disease and obtained encouraging results. However, bibliographic networks are heterogeneous in reality, such that there are multiple types of objects (\emph{e.g.}, venues, papers, topics) and multiple links among these objects. The prediction of various types of relationships in different social networks deserves further study, as it may be conducive to other applications, such as community detection \cite{yu2017datacom}, influence analysis, and link recommendation \cite{xia2014tetcMVCWalker}.

Above all, the focus of relationship prediction is whether a social tie will appear in the future. However, it seems to be a more interesting problem to identify which point in time the relationship will be built or has been established. Sun et al. \cite{sun2012will} extracted topological features from heterogeneous networks (like DBLP bibliographic network) to formulate a generalized linear model for relationship building time prediction. Meeder et al. \cite{meeder2011we} proposed a time-stamping method for the Twitter social network which took a single static snapshot of network edges and user account creation time, to accurately infer when the links were formed. It is helpful for relationship evolution through learning about the relationship building time, thus we can study how the relationship changes from the establishing time as well as the strength of relationships as time progresses.

\textit{3) Relationship Evolution}

A social relationship is subject to constant evolution due to frequent changes in individual activities and interactions. The emergence of a data-driven CSS provides dynamic and continuous human interactions reflecting the evolution process of relationships between individuals. However, tracing a relationship's temporal evolution appears to require multiple observations of the social network over time. These expensive and repeated crawls are only able to answer questions from observation to observation, but not what happened before or between network snapshots. Therefore, numerous researchers concentrate on the influential factors of the process of relationship evolution and ignore the entire relationship lifecycle.

Some studies have investigated what effects specific factors have on social relationships as time evolves. These include online social networks, user behaviors or activities, and mutual interactions. Vitak et al. \cite{vitak2014facebook} demonstrated that online social networks, such as Facebook, helped to maintain friendship among users who are geographically dispersed and communication-restricted. With respect to the influence of user behaviors on relationship evolution, Dong et al. \cite{dong2011modeling} described a Markov jump process model to capture the co-evolution of relationships and behaviors in a student dormitory. The dataset was collected based on monthly surveys and locations/proximities traced by cell phones for a long period, so as to demonstrate the rate of visiting places and friends. For mutual interactions, Yu et al. \cite{yu2013understanding} explored social relationship evolution based on MIT mobile phone data by using Heider's Social Balance Theory, and verified that the reciprocality and transitivity play an important role in social relationship evolution.

Social relationships are usually reflected by user interactions \cite{xia2016tmcpis, xia2014tpdsreplica}. This leads to some investigations on revealing the disciplines of individual interactions, by using real-world sensing or online social network data. Cattuto et al. \cite{cattuto2010dynamics} presented a scalable experimental framework for gathering real-time data by resolving face-to-face social interactions with tunable spatial and temporal granularities. The relationship between individuals is incarnated through their interactions, and the frequency of communication may represent the strength of social relationships to some degree. Thus, discovering the patterns of interaction contribute to an understanding of the process of relationship evolution.

\subsection{\textbf{Collectives}}

\begin{wrapfigure}{R}{0.5\textwidth}
  \begin{center}
    \includegraphics[width=3.2in]{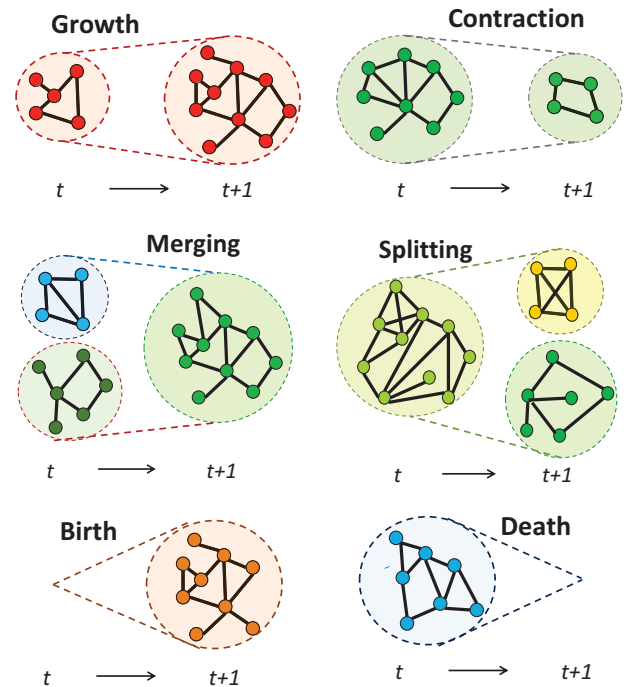}
  \end{center}
\caption{Types of community evolution.}
\label{codynamics}
\end{wrapfigure}

In this section, we will discuss how to analyze human behaviors at a collective scale. Human behavior analysis involves a great deal of human-to-human relationships, human-to-community relationships, and community-to-community relationships. The existence of communities in a social network is undoubted \cite{xia2014community}. When we try to understand human behavior, it is necessary to analyze this issue from a collective perspective. Exploring communities in mining human dynamics is essential and of great importance and it has been studied for a long time. First of all, people are more likely to interact with others who have similar interests and they will form a group. We are interested in identifying these groups for specific purposes, for example, advertisement recommendation. Secondly, the community provides a global view of human dynamics \cite{yu2019tccsteam, yu2019csrteam}. Collective behavior is more stable, whereas individual behavior is noisy and easy to change. Finally, some human behavior can only be observed from a community setting and not from an individual level.

Usually, the human community can be defined in a strict sense as a group of individuals who are more densely connected within a group than others \cite{browet2017incompatibility}. For example, people with similar social properties (i.e. expertise, interest, background, etc.) are more likely to be friends. Indeed, social communities are ubiquitous, arising in every type of social organization: universities, political and business organizations, clubs, and even in virtual communities such as Facebook groups. One of the most famous studies of the human community is the Zachary karate club network \cite{zachary1977information}, which has been studied by many researchers \cite{yang2015defining, newman2012communities}. This community is constructed with the data collected by observing $34$ members of Zachary karate club for more than $2$ years and considering friendship between members. Because of the disagreement between the leaders, individuals split into two different communities.

\textit{1) Community Detection}

Community detection may reveal important functional information about human interactions. In social networks, technologies about community detection can be used in identifying the most influential people, recommending potential collaborators, analyzing social relationships, and so on. As a result, the problem of identification of communities has been the focus for many years \cite{kheirkhahzadeh2016efficient}. Depending on the underlying methodological rule as well as the different features of communities, we introduce four common community detection methods:

\textit{a. Modularity-based Methods }

Many methods are founded based on the principle of optimizing the graph-based measure of community quality, for example, modularity \cite{newman2004fast}, which is one of the most popular quality function. The methods in this classification aim to design optimal graph-based measures to detect community quality. However, maximizing $Q$ is an NP-hard problem. Many approaches have been proposed to solve this problem including greedy agglomeration \cite{newman2004detecting}, mathematical programming \cite{agarwal2008modularity}, sampling techniques \cite{sales2007extracting} etc. One of the most famous algorithms devised to maximize modularity was a greedy method by Newman \cite{newman2004detecting}. He proposed a fast clustering method that uses a greedy strategy to get a maxim $\Delta Q$ by merging pairs of nodes iteratively. Rees et al. \cite{rees2012overlapping} proposed an algorithm which first builds an ego network as a friendship community for each node. Then it compares all the ego networks to form a whole picture of the network with identified communities.

\textit{b. Divisive Algorithms}

These methods identify and remove nodes or edges among communities via measures such as betweenness centrality. An edge has high betweenness if more short paths go through it. One can detect communities through a process of ranking all the edges based on betweenness. Girvan and Newman \cite{girvan2002community} proposed a GN algorithm, which is a typical divisive algorithm. The GN algorithm progressively computes betweenness for all edges and removes the edges of a network with the highest score. They also considered three other definitions including geodesic edge betweenness, random-walk edge betweenness, and current-flow edge betweenness.

\textit{c. Social Interaction-based methods}

Previous algorithms mainly focus on topological features with the following assumptions: the relationships within the community are fully known; the relationships among nodes are of a single type. Problems occur when we try to identify a hidden community with missing links or with multi relationships. In the light of above weaknesses, Chen et al. \cite{chen2014community} proposed a new social interaction-based approach to identify Facebook community, where social interaction refers to users interacting behaviors with others, such as posting, reading, replying, and so on. Their approach extended previous structure-based community detection methods with social interaction, which allows them to consider more features.

\textit{d. Overlap Community Detection}

Community detection methods mentioned above can only discover non-overlapping community structure ignoring the fact that communities may be overlapped. Overlapping community detection becomes another crucial task in human collective behavior analyzing \cite{amelio2014overlapping}. The existing algorithms for detecting overlapping community can be summarized in three kinds:
\begin{itemize}
\item Derive from none-overlap community detection algorithm and form new modularity based on a probabilistic model in order to detect overlapping community \cite{newman2007mixture}.

\item Take advantage of a clique to discover overlapping community, which is based on the assumption that a community consists of overlapping groups of fully connected sub-graphs and communities can be detected by searching for adjacent cliques. For example, Palla et al. \cite{palla2005uncovering} introduced an approach to analyze the main statistical features of the interwoven sets of overlapping communities based on the Clique Percolation Method (CPM).

\item Use link partition rather than note partition to detect overlapping community \cite{wang2010discovering, ahn2010link}. The main idea underlying this method is that a note can only belong to one community but it may have several edges, which means that a note can be assigned to multiple communities as long as its edges can be assigned to multiple communities.

    \end{itemize}

\textit{2) Community Evolution}

Community detection methods introduced above assume that networks are static, which means that the nodes and links among the communities are stable and do not change over time. In reality, human society is a temporal social network (TSN). Because of the frequent changes in the behavior and communication patterns, the associated human communities are subject to constant evolution. Having a comprehensive knowledge of community dynamic analysis, one may predict the future of the community and then manage it properly. Thus, it is essential for a deeper understanding of the development and self-optimization of human society. Community evolution is a sequence of community changes succeeding each other in the consecutive time windows within a social network \cite{brodka2013ged}. In \cite{barabasi2002evolution}, it proposed six types of community dynamic events, as shown in Fig.~\ref{codynamics}:
\begin{itemize}
\item Growth: A community grows when new nodes have joined in the community.

\item Contraction: A community shrinks when some nodes have left the community.

\item Merging: Merging happens when a new community has been created by merging several other communities.

\item Splitting: Splitting happens when an old community split into two or more communities in the next time window.

\item Birth: Birth occurs when a new community, which does not exist in the previous time window, appears in the next time window.

\item Death: Opposite to Birth, Death means that a community ends its life and will not appear in the next time window.
\end{itemize}

Several methods have been proposed to analyze the community evolution. Palla et al. \cite{palla2007quantifying} developed an algorithm based on clique percolation in order to uncover basic rules characterizing community evolution. They focused on scientific collaboration networks and mobile phone call networks. They found that large groups persist for a longer time if they could dynamically alter their membership, indicating that an ability to change the group composition results in better adaptability. Piotr et al. \cite{brodka2013ged} proposed Group Evolution Discovery (GED) to explore the evolution of the social groups. The GED method used not only the size and equivalence of communities' members, but it also took their position and importance within the group into consideration in order to analyze what happened within the group. Sun et al. \cite{sun2013social} took advantage of scientific community dynamics to study the birth and decline of disciplines. They proposed an agent-based model in which the evolution of disciplines is affected mainly by social interactions among scientists. They found out that disciplines emerge from splitting and merging of social communities in a collaboration network.

\textit{3) Community Behaviors}

The definition of collective behaviors has been broadened during the past century. However, the core concept remains the same, which considers collective behavior as ``an adaptive response to new or ambiguous condition" \cite{turner1957collective}. These behaviors are often unplanned and spontaneous. For instance, individuals go out for shopping, standing in line for a new product release, and rush away after the occurrence of accidents or emergency situations. We need to analyze these collective behaviors in order to better understand the potential reasons underlying a phenomenon \cite{miller2013introduction}. Thanks to the fast development of technologies, we are now able to get massive real-time data associated with collective dynamics. Collective behavior analysis can be performed by analyzing individuals' behavior \cite{song2010limits}. That is, one may either divide collective behavior into small groups and analyze them accordingly or analyze the population as a whole. The first situation will result in an expected behavior by putting the divided analysis together. The latter approach is to some extent the same as analyzing an individual, with the difference that links are now considered from a community perspective.

Among many aspects of collective behavior analysis, we are interested in the following topics, collective decision-making, cooperations and contagions among collectives, human mobility, and collective behavior prediction.

\textit{a. Collective decision-making}

Individuals with similar attributes or purposes are often deemed as being in the same communities. Although members in the same community share some common characteristics, they may hold different opinions toward different alternatives. Therefore, when collectives have to choose among diverse choices, it is fairly difficult to reach the common decisions which are accepted by all the group members. As a result, how to choose the suitable choice for the collectives and taking members' different demands as many as impossible into consideration remain to be an open area for researchers. This is where computational social choice comes into play.

Social choice theory is concerned with the analysis of collective decision making \cite{nunez2015social}. The traditional social choice theory addresses various kinds of questions, such as how to choose among several alternatives or how to rank those choices by mostly using mathematical and economic theories. Yet with the advancements and aids of computer science, the field of research problems are greatly extended now referred as computational social choice \cite{chevaleyre2007short}. Computational social choice is an interdisciplinary research area that involves mathematics, social choice, economics, political science, and computer science. There are several main research directions which can be used to study human dynamics that computational social choice concerns about:
\begin{itemize}
\item Preference aggregation.
   Different individuals have different preferences toward the same things. Gathering these preferences and mapping them into the set of collective preferences are the commonly studied problems in this area which is more inclined to mathematical presentations. Using logical or graphical compact representation languages to present preferences with combinatorial structures and selecting appropriate functions to map those preferences to single or nonempty alternatives are the most frequently used methods in preference aggregation.

   The applications that involve human dynamics using this area of methods are, for instance, community detection and peer review process. In \cite{kanawati2014seed}, it used a seed-centric algorithm that borrows preference aggregation strategies to discover communities, and this kind of method is a new trend in community detection. In the peer-review process \cite{roos2014algorithms}, it used preference aggregation rules to integrate the diverse aspects of papers and to rank those papers.
\item Ranking and the Internet searching engine.
   Ranking systems are developed to help individuals to choose the most appropriate choices for themselves according to their own requirements. The typical PageRank algorithm has been widely used both in ranking systems and the Internet searching engines \cite{page1999pagerank}. A lot of research extends PageRank algorithms due to specific conditions taking individuals' various preferences into considerations to develop new searching engines \cite{nikolakopoulos2013ncdawarerank}.

   Despite the above applications on ranking and searching methods, there also exist other kinds of utilizations that involve with human dynamics by using this area of methods, for instance, recommendation systems \cite{xia2017isjrec} and evaluation of researchers' impacts. In \cite{albanese2013multimedia, otegi2014personalised}, they used PageRank or extended PageRank algorithms to rank products, journals, friends, and potential cooperators by using data collected from online social networks, and then recommended it to the users according to the rank lists.
\item Resource allocation.
   How to divide a cake seems to be a simple question that has been studied by scientists for a long time. It requires distributing definite resources to a set of individuals. Agents communicate with each other to express their preferences and requirements in order to get the most suitable sources under the precondition that the benefits of other people are not harmed. There are two fundamental criteria that should be considered when allocating resources, which are fairness and efficiency. Based on these two crucial factors, several allocating mechanisms are explored.

   The application that involves human dynamics using this area methods is, for instance, ranking authors' sequence. Schall \cite{schall2013measuring} indicated that different author's goals such as career requirements for promotion to rank the sequence should be considered. Ackerman et al. \cite{ackerman2014authorship} proposed a game-theoretic model to study the allocation of credit to authors, and the results show that alphabetical ordering can lead to higher research quality while ordering by contribution results in a denser collaboration network and more publications.
\item Voting.
   Voting is the most widely used method to reach uniform decisions when collectives meet a group of alternatives, which is not limited to single-choice problems. The scenarios of voting are also diverse such as political elections, items' selection, and performance's estimation, etc. In the voting process, due to different circumstances and requirements, researchers proposed many specified rules such as plurality rule, Borda rule, Condorcet-consistent rule.

   The applications that involve human dynamics using this area methods are, for instance, recommendation system, political election, and rating system. In \cite{popescu2013group}, it proposed an online group recommendation system using voting strategies to let the system have a better understanding of users.
   \end{itemize}

\textit{b. Cooperation and Altruism}

A typical type of relationship among individuals within the same communities is cooperation. Cooperation has always been the core problem that involves with interests of individuals and collectives, and attracts researchers from diverse areas to investigate. Cooperation exists universally in our daily lives, however, human beings are both rational and emotional when making choices related to their own benefits, sometimes they are selfish, and sometimes they would choose to cooperate with others even bare costs instead of maximizing their own profits over the whole society. These complex cooperation and altruism phenomena attract scientists to discover their inner mechanisms and under what circumstances individuals would consider cooperating with others instead of being purely self-centered.

Before the era which restricted to the development of technologies, it is hard to record and access human beings' daily activities. Therefore, most researchers had to study cooperation issues based on simulation experiments. In 1992, Nowak et al. \cite{nowak1992evolutionary} introduced a spatial structure that enables cooperators to form clusters and protect them from being exploited by the defectors using a square lattice network. Standing on the shoulders of them, the subsequent works have investigated the spatial structure's impacts on the evolution of cooperation and proposed various strategies to promote cooperation. Nonetheless, Hauert et al. \cite{hauert2004spatial} found evidence that spatial structure is not always beneficial for promoting cooperation, thus recent researchers begin to investigate evolutionary games on complex networks. Researchers also explored the impacts of network structures on cooperation, which are scale-free network and small world network \cite{ichinose2017invasion}, \cite{xu2016understanding}. Vilone et al. \cite{vilone2012social} proposed strategies from various areas combined such as the voter model and threshold rules with using different social dilemmas to promote cooperation based on comparing simulation results.

An important mechanism that affects the evolution of cooperation is altruism or reciprocity \cite{croson1998theories}. People sometimes are willing to help others even under the circumstances that they have to pay the cost due to complex reasons, which we call altruism. There are several specific mechanisms that affect cooperation among individuals which are kin selection \cite{queller2016kin}, direct reciprocity, indirect reciprocity, network reciprocity, and group selection.

With the development of technologies, more and more people are participated in online social networks and left a huge amount of data that can be recorded and traced about themselves and their interactions with others. By analyzing these data, it can provide us looking into the cooperation problem from a whole different perspective. Due to the privacy and inconvenience of collecting data on relations about humans' daily behaviors, more and more researchers focus on the study of cooperation issues using scholarly/academic data \cite{khan2017survey}, i.e., scientific collaborations \cite{Kong2019Academic, Wang2017sclosure}.

There are two general directions to investigate the mechanism of scientific collaboration using scientific data sources. From the perspective of network analyses, Newman \cite{newman2001structure} explored the structure of scientific collaboration or coauthorship networks, in which scientists are connected if they have coauthored papers by using data collected from various databases. Their works indicated that scientific collaboration networks tend to form ``small world'' \cite{newman2004coauthorship} and discovered a variety of attributes about collaboration patterns. Following his pioneering works, researchers then used diverse databases on various disciplines to establish collaboration networks and investigate the impacts of the structures on the collaboration patterns. Another type of network that scientists explore using scholarly data is the citation network. Researchers in \cite{xia2016bibliographic} found that either in the coauthorship or citation network, the nodes and structure can have impacts on each other, and also the comprehensiveness of scholarly data sources can have a significant influence on the network structures and its attributes.

Another research perspective focuses on analyzing the data. The records of scholarly data contain information on authors, keywords, title, cite papers and publish information, etc. Taking full advantage of these data, we can find answers to many problems. Bales et al. \cite{bales2014associating} also find that the more coauthors no matter from inter-discipline or multidiscipline that collaborate in one paper, the quality of it and the chance of being accepted are higher compared to those whose coauthors are from mono-discipline. Ozel et al. \cite{ozel2014co} indicated that the coauthorship pair distribution can be varied by gender, discipline, and countries.

Before tackling all the above questions that emerge after finishing the papers, it is also important to find the comparative researchers to cooperate with for people who want to start writing papers. Schall \cite{schall2014multi} analyzes the personal files, research interests, and their relationships with other researchers, to find the suitable cooperators. Together with using recommend algorithms that based on typical PageRank or k-shell algorithms, it can provide a list of potential coauthors that researchers can cooperate with, which can maximize the scientific performances \cite{xia2016scientific, Kong2020tkdd}. Another kind of databases such as actors' cooperation in TV shows or films also can be traced and used to study cooperation phenomenon in entertainment industries.

\textit{c. Contagion}

Another problem that involves the vast majority is social contagion, which utilizes the relationships among individuals to diffuse. Any goods, emotions, behaviors, or viruses can propagate through it. Each of these diffusion processes through the population based on some specific mechanisms by exploiting interactions among individuals. Scientists with diverse backgrounds have been studying these problems for a long time in order to understand these rules and take full advantage of it.

In the beginning, researchers study these propagation problems mainly on the virus epidemic. Restricted to difficulties in keeping track of human mobility and people they connected, most methods are based on simulation models. In the classic infection models, they assume individuals have three conditions, which are susceptible, infected, and recovered. It assumes that individuals get connected with anyone in the population with the same probability. There exist three typical models which are SI model, SIR model, and SIS model \cite{anderson1991infectious}. The basic differences between them among individuals' condition after being infected which in SI, individuals stay in infected condition permanently; individuals can recover from the infected condition with constant velocity in SIR; individuals can be back to the susceptible station again after recovering from the infected condition in SIS.

On the basis of the above models, and with the help of the more and more detailed data which describe the process of virus contagion, researchers proposed a variety of more accurate methods to study the diffusion of virus compared to those merely based on simulation and mathematical deductions. Balcan et al. \cite{balcan2009multiscale} found diverse important factors that should be considered in spatial studies of human diseases, such as household, community attributes and etc. Following the works in \cite{balcan2010modeling}, it presented the model that integrates socio-demographic and population mobility data in a spatially structured stochastic disease approach to simulate the spread of epidemics at the worldwide scale. Among the above factors that should be considered to study the virus epidemic, the mobility data of individuals are of importance to supervise and control the contagion of the virus. Due to the difficulty of acquiring people's movement data, Tozzoni et al. \cite{tizzoni2014use} used human mobility proxies to model epidemics and evaluated different proxies' accuracy. Yet in addition to the virus that makes people unhealthy, it also exists in the virtual world that may lead to the whole paralysis of systems.

Earlier research proposed models \cite{dodds2004universal} to study the patterns and their universal behaviors of information or rumors, products, and even behaviors' diffusion by using general contagion models. However, different kinds of things that diffuse through population should be explored via different methods considering their diverse application scenarios in reality. The introduction of diverse methods that coact with detailed data to study diffusion by different classifications will be introduced in the following, where we mainly focus on information and behavior diffusion \cite{bozorgi2016incim}.

In order to study information diffusion, Goel et al. \cite{goel2013structural} analyzed events diffusion process which includes news, stories, videos, images, and petitions on Twitter and discovered different kinds of events' diffusing patterns, which are characterized by different types of structural virality. Most of these methods hypothesize the network structures are homogeneous, while humans are connected via different types of relationships in the real world. Gui et al. \cite{gui2014modeling} proposed an information diffusion model in multi-relational networks, by distinguishing the power in passing information around for different types of relationships. In addition, it applied the model to DBLP network and APS network, and experimentally demonstrated the effectiveness of their methods compared with single-relational diffusion models. Except for the network topologies, most of the methods also assumed that people can only be influenced by other members of the network \cite{tang2013confluence}, where information spreads because of informational cascades. However, most observed spreading processes in online social networks do not rely solely on social influence. The closed-world assumption is proven incorrect in recent work on Twitter done by Myers et al. \cite{myers2012information} in which authors observe that information tends to jump across the network. Consequently, they provide a model capable of quantifying the level of external exposure and influence using hazard functions.

In 2010, Demon Centola conducted an online experiment to prove that information diffusion is different from behavior spreading, which individual adoption was much more likely when participants received social reinforcement from multiple neighbors in the social network \cite{centola2010spread}. It indicated that information diffuses through weak ties, and strong ties can enhance behavior diffusion. As a matter of fact, before this experimental works, some researchers used threshold models \cite{granovetter1978threshold} to describe the process of behavior diffusion which conformed to Centola's results. Yet other works proposed methods on behavior diffusion that have been used to study information diffusion, they are beneficial to some extent but not suitable to the reality. Currently, pure or hybrid models are used to study the diffusion problems which considering individuals' material benefits \cite{vilone2012social}, the psychology of conformity with the majority or peer influence to simulate the process of behavior diffusion. And the results indicate that the performances of those models can be improved by varying parameters and also associate with network topologies. Due to the disabilities in tracing various behaviors' diffusion process, most of the studies are inclined to use simulate experiments.

In addition to research on information and behavior diffusion, there also exist other kinds of contagion issues. Researchers investigated obesity, smoking, crimes, or emotions' diffusion patterns \cite{kramer2014experimental}. They discovered that various factors can affect those behaviors, such as climate, seasons changing, or opinions from people around them and etc by using online social network data. These works that explore social problems by using computational methods can have significant impacts on the evolution of human society.

\textit{d. Human Mobility}

Uncovering human mobility behavior is crucial for future predicting and controlling spatially embedded events such as disease control \cite{miller2017mathematical} and traffic forecasting \cite{kong2016urban}. Individuals have highly different mobility patterns \cite{zhong2016variability}, while relative analysis shows regular displacement distributions ranging from power laws \cite{gallotti2016stochastic} to exponential laws. By human mobility analyzing, we aim to study the collective mobility behavior of a group of individuals. Compared with individual mobility traits, it focuses on the whole aspect of human moving patterns and rules. For example, it may refer to the study of the collective behavior of taxi drivers or passengers of a city based on their digital footprints \cite{kong2017time}.

Most research on human mobility is performed from location-based devices such as cell phones, portable computers, and GPS localizers. These devices record digital footprints of human activities which can reflect the social interactions of a community. Taking advantage of techniques from many fields, we could obtain a better understanding of the underlying dynamics of individuals or communities by extracting communal behaviors from these data \cite{ratti2011social}.

Extracting hotspot, which is the most frequently visiting location in a city, would be a meaningful research area of analyzing human mobility behavior. There have been extensive studies on detecting significant places using GPS trajectories from personal devices. In \cite{d2017detection}, it presented an expert system for detecting traffic congestion and incidents from real-time GPS data collected from GPS trackers or drivers¡¯ smartphones. Aside from using mobile phone data, there are some other works using digital footprints from other devices for detecting interesting places. One representative study is the GeoLife project \cite{zheng2010geolife} which uses data collected from the raw GPS trajectories over various transportation modes of $167$ users over $3$ years. Another well-known work is Dartmouth's CenceMe project \cite{miluzzo2008sensing} in which the authors design intelligent mobile sensor networks which may sense nearby friends and their current activity.

Other research on human mobility is performed from social media check-in data. Yang et al. \cite{yang2017indigenization} investigated the different mobility patterns between natives and non-natives. They performed intensive analyses on a data set including millions of geographical check-ins in five large cities in China from a location-based social network service. And they found distinguishable mobility patterns of natives and non-natives. In \cite{beiro2016predicting}, it proposed a hybrid model of human mobility that integrates a large-scale publicly available dataset from a popular photo-sharing system with the classical gravity model, under a stacked regression procedure.

\textit{e. Collective Behavior Prediction}

Collective behavior refers to behaviors of individuals in the space at the same time, but it is not just the aggregation of individual behaviors. With easy access to massive data associated with human collective behavior, we may investigate its power at predicting real-world outcomes. Studies have shown that community can indeed be used to make quantitative predictions on human collective behavior, especially on artificial market \cite{wang2016towards}. This digital information generally involves the comprehensive detail about human preferences, and if large enough and properly designed, they will be more accurate than other techniques for information suggestion, such as surveys and opinions polls.

For collective behavior prediction, researchers mainly take advantage of people's digital data to analyze and predict what will happen in the near future in order to offer more comfortable and personal services. They mainly use regression analysis to study what will happen or what is most relevant to given information based on collective data. For example, much research has been done on behavioral economics with social media data \cite{pineiro2017influence}. Li et al. \cite{li2016tensor} introduced a tensor-based information framework to predict stock movements, which models the complex investor information environment with tensors.

Other recent research also indicates that consuming behavior may be unpredictable but relative indicators can be detected from online social media (blogs, information sharing, etc.) to predict changes in various economic and commercial trends. For example, Salehan et al. \cite{salehan2016predicting} investigated the effect of review sentiment on readership and helpfulness of online consumer reviews. Furthermore, Beauchamp \cite{beauchamp2017predicting} combined 1,200 state-level polls during the 2012 presidential campaign with over 100 million state-located political tweets to predict opinion polls using a new linear regularization feature-selection method.

\begin{table*}[t]
\centering
\caption{Comparision of different data resources.}
\label{tab:data}
\begin{tabular}{|m{3cm}<{\centering}|m{3cm}<{\centering}|m{3cm}<{\centering}|m{4cm}<{\centering}|}                                               \hline
Data Resource    & Application Area   & Devices  & Examples               \\ \hline
Web Resources    & Text, web, multimedia, network analytics     & API interfaces; web crawlers     & Facebook; Twitter; Email; logs etc. \cite{park2016women, goel2010predicting}         \\ \hline
Sensing Data     & Mobile analytics, Urban computing      & Mobile apps; Sensors; RFID; wearable devices and etc.  &  GPS data; health condition etc. \cite{cattuto2010dynamics, tizzoni2014use, lu2013approaching}            \\ \hline
Data from Self-designed Experiments        & Structured data analytics   & Self-designed experiments online or offline, online experiments platforms  & Scientific experiments data \cite{chittaranjan2011s, vitak2014facebook, centola2010spread}              \\ \hline

\end{tabular}
\end{table*}

\section{Research Methodology}

%\begin{wrapfigure}{R}{0.5\textwidth}
%  \begin{center}
%    \includegraphics[width=3.2in]{codynamics.eps}
%  \end{center}
%\caption{Types of community evolution.}
%\label{codynamics}
%\end{wrapfigure}

%
%\begin{figure*}
%\centering
%\includegraphics[width=7in]{dataAnalysis.eps}
%\caption{Data analysis methods.}
%\label{dataanalysis} % Give a unique label
%\end{figure*}

\begin{wrapfigure}{R}{0.5\textwidth}
  \begin{center}
    \includegraphics[width=2.6in]{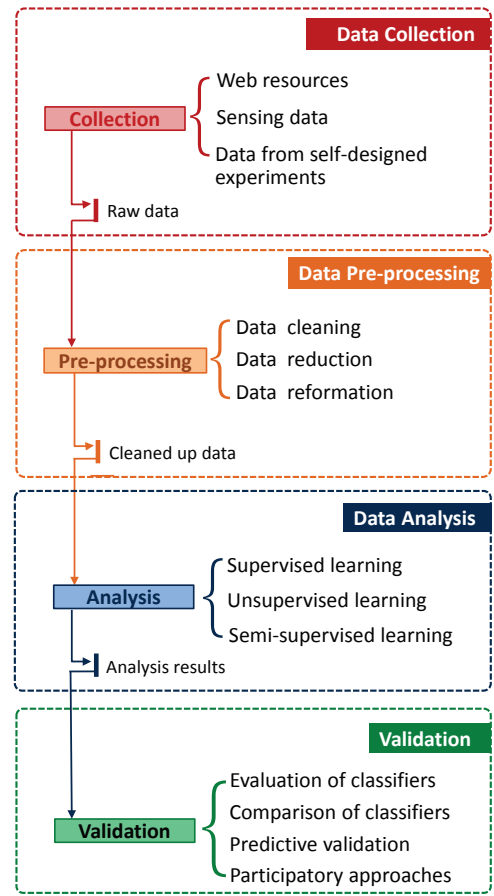}
  \end{center}
\caption{The process of solving CSS issues.}
\label{fig4}
\end{wrapfigure}

All the research begins with meaningful issues that are raised to solve particular problems in different domains. The range of data-based problems is expanding all the time with the development of new technology and the accessibility of new data \cite{Liu2019Retrieval, liu2019sis}. Due to the explosion of information, using computational methods and combining the traditional approaches to study social science can lead us to different directions to explain social phenomena. The research methods in CSS follow several general steps, which are data collection, data pre-processing, data analysis, data validation, and the detailed contents of each step are shown in Fig.~\ref{fig4}.

\subsection{\textbf{Data Collection}}

Once the research topics are confirmed, the investigation focuses preliminarily on the selection and collection of data such as the content or the scale of the data. Data plays a crucial role in exploring social issues, meanwhile, choosing the appropriate data sources can greatly improve efficiency and save us a lot of extra manual operations. There are several kinds of data sources that can be taken advantage of when facing social problems.

\textit{1) Web Resources}

The first data aggregation place is the World Wide Web. Nowadays accessing the Internet has become more and more convenient, and individuals can connect to it almost anytime anywhere. A huge amount of digital records about ourselves are traceable online and can be used for further research. Data from these sources can be used to mine user profiles, users' online social life, organization structures, and social relationships. Being aware of where to get the data, the next step is how to get. Due to individuals' privacy issues, some related data cannot be accessed by the public which can only be acquired by authorized organizations. Apart from those data, we will introduce the most commonly used methods of collecting public data.

The most fundamental way of collecting online data is through API interfaces that are provided by the websites. While not all websites offer those API interfaces, so how to collect data without them? Web crawlers are really required under this circumstance. Typically web crawlers tend to use the Python language for its advantages in acquiring data. Users can write their own web crawlers to get the exact data as they command. In \cite{park2016women, goel2010predicting}, the data are all collected by web crawlers on Facebook and search engines.

\textit{2) Sensing Data}

Targeting at solving different kinds of research issues, another way of collecting data is through sensors, RFID, mobile devices and etc. Along with the advancements of wireless sensing technologies, wireless, wearable, and mobile devices can be used to collect data. Infrastructure-bound sensor devices which mainly provide us data about the environment, as well as indoor or outdoor activities of individuals or groups. Mobile and wearable sensor devices which are user centric and catch data about user location, user movements, and interpersonal interactions.

The data collected from the above devices can shed light on urban computing issues which contain traffic problems, environment monitoring, human mobility issues, health problems and etc. \cite{zheng2014urban}. Tizzoni et al. \cite{tizzoni2014use} analyzed data collected from mobile devices to study the impacts of human mobility on contagion issues. Through collecting mobile phone call records, Xin et al. \cite{lu2013approaching} analyzed individuals' travel patterns in Cote d'Ivoire. Cattuto et al. \cite{cattuto2010dynamics} used RFID tag information to explore the interactions between individuals.

\textit{3) Data from Self-designed Experiments}

Applying the above general methods of data collection can provide us with a very plentiful amount of data. In spite of the acquisition of data that can meet the command of research topics, when facing problems that cannot find the appropriate data sources, how to get the targeting data for those problems is still need to be considered. In traditional social science, sociologists issue questionnaires that contain questions they need to know about individuals and then get the relevant data, however, these works rely heavily on human labors.

Currently, with the aid of both computer technologies and mathematical theories, most of these questionnaires can be done online and the acquired data are processed by specific applications. And there have been various online experiment platforms like Mechanical Turk, that can help scientists conducting their own experiments and get the data they need.

Due to the difficulty in recording human behavior data, Centola \cite{centola2010spread} designed an online experiment to collect data on innovation behavior and analyzed its mechanisms. Despite designing experiments, issuing questionnaires is also an effective way of collecting the appropriate data as shown in \cite{chittaranjan2011s, vitak2014facebook}.

Recently the above methods of data collection are the most commonly used approaches. We compare the difference of each method as shown in Table~\ref{tab:data}. However, results relying on either of these data resources have limitations. In future studies, the incorporation of these three sources should be considered in order to get overall data and have a better understanding of research issues. With the development of technology, data procurement becomes more facility but also faces difficulties. While the vast online and offline datasets produced by human beings have the potential for a good result of the analysis, there is no guarantee that we can deal with so rapid increase in human data. And the large-scale data inevitability leads to longer processing time. The intellectual property rights and privacy make hard access to the data sets, which are the most prominent challenges of gathering the data.

\subsection{\textbf{Data Pre-processing}}
With the aid of analyzing methods on the huge amount of data, we can discover various patterns or mechanisms to reveal the nature of our society. Due to the heterogeneity of network structures, the sources of data could be diverse. Therefore, the structure of raw data collected from different sources can be heterogeneous. Using these raw data not only wastes lots of human labors, but also affects the analyzing results. Before analyzing the data, it needs to be processed into recognizable formats that require being cleaned and reorganized. Pre-processing the raw data can greatly improve the quality of data and significantly enhance the efficiency of the analyzing process. The procedure of data pre-processing is necessary and the following are the main steps.

According to the above analysis of data collection, it is obvious that the collected data, which is heterogeneous and very messy, cannot be used immediately. Therefore, before data analysis, pre-processing the raw data can greatly improve the quality of data and significantly enhance the efficiency of an analyzing process. The procedure of data pre-processing is necessary and the following are the main steps.

\begin{itemize}
\item Data cleaning. Initially, the collected data may be incomplete, inconsistent, noisy, redundant. To eliminate its negative effects on the following analysis procedure, it is essential to clean the datasets at the beginning. There are several approaches to handle the raw data. For the incomplete tuples, we can either choose to ignore them or fill in the missing values according to specific rules such as removing missing values related instances, replacing missing values with the most common constant, and replacing missing ones with values estimated by other features of the same instance. While when the datasets are noisy, we can apply several smoothing techniques such as binning, regression, outlier analysis and etc. In \cite{crandall2010inferring}, it filtered photos with imprecise geo-tags and/or missing timestamps to improve the quality of data. In addition, data that cannot meet the requirements is also deleted in \cite{de2013predicting}.

\item Data reduction. It is undeniable that we are entering the big data era. The data scale nowadays we are dealing with is huge which may slow down the processing speed and need a lot of extra labors. Data reduction can be used to obtain a reduced representation of original data sets and still get the almost same analytical results. Generally, there are two directions to reduce the data scale which are dimensionality reduction and numerosity reduction.

    Dimensionality reduction methods mainly focus on transforming the original data into a smaller space, such as wavelet transforms, principal components analysis, and attribute subset selection. Singular-value decomposition method is used to reduce the dimensionality of the data set in \cite{gorinova2016predicting}. Feature extraction is also used in dimensionality reduction \cite{chen2013friendship}. While numerosity reduction approaches use different representation methods to store data and decrease the volume, which includes parametric and nonparametric methods. In addition, sampling methods are used to reduce the size of data and improve the processing efficiency \cite{zhao2013emergence, gonzalez2008understanding}.

\item Data reformation. The last step is the data reformatting. A single whole database sometimes is needed to be divided into smaller component units to do further analysis. The features of the data also require aggregation or discretization analysis, providing new features which can perform the research task better. Strategies of data reformation include smoothing, attribute construction, aggregation, normalization, discretization, and concept hierarchy generation for nominal data. As in \cite{min2013mining, dong2011modeling}, different classifications of data resources are combined together to analyze human relationships.
\end{itemize}

\begin{figure*}
\centering
\includegraphics[width=7in]{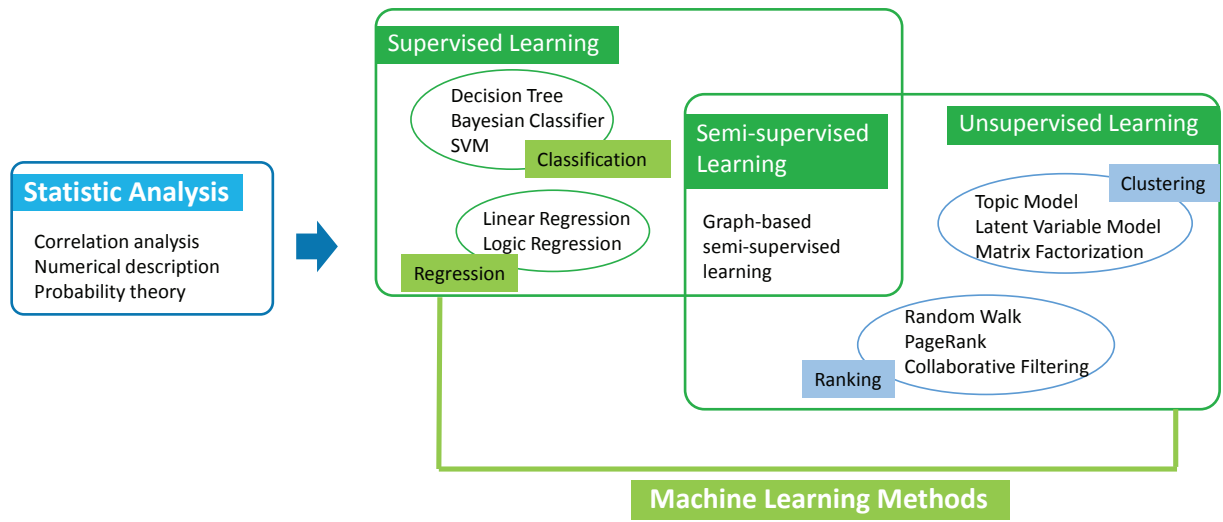}
\caption{Data analysis methods.}
\label{dataanalysis} % Give a unique label
\end{figure*}

Another problem that needs to be considered in the data pre-processing procedure is the multi-source data fusion circumstance because using single data input maybe not adequate for some situations. Therefore, utilizing multi-source data has become increasingly important and some achievements have been made. For instance, researchers in the information system area try to solve the multi-source data fusion problem from the perspective of information entropy \cite{Li2017fusion}. Liu et al. \cite{Liu2019general} propose a general multi-source data fusion framework that solves the heterogeneity and semantic conflict in the big data environment. In internet of things \cite{bin2020property, Raffaele2016body}, scientometrics \cite{Xu2017sciento}, network science\cite{min2020Mechanical} and material science areas \cite{zhou2019Material}, researchers have designed diverse multi-source data fusion methods which aim at their own discipline characteristics and issues. In the future, the multi-source data fusion problem will still be challenging under the circumstances of the rapid growth of data and the interactions between offline and online.

\subsection{ \textbf{Data Analysis}}
Once the data has been cleaned up, the next stage is data analysis, which deeply mines various features of the data and found regular results to solve the research questions. This is one of the most crucial stages of the research process. The data collected from Internet Social network and interaction services, infrastructure-bound sensor devices and mobile and wearable sensor devices can include text, images, graphs, or any other types of form after pre-processing. So how to extract useful information from these diversiform forms of data set? Some frequently-used and well-performing analysis approaches are shown in Fig.~\ref{dataanalysis}, and the specific algorithms are introduced as follows.

Traditional data analysis means to use previous statistical methods to analyze massive first-hand data to concentrate, extract, and refine useful data from raw data. Statistical analysis is based on statistical theory, which is a branch of applied mathematics. In statistical theory, researchers often model randomness and uncertainty with Probability Theory. By analyzing the statistical features of data such as mean, variance, coefficient, entropy, maximum/minimum value, and mathematical distribution, we can identify the inherent laws of the subject matter, so as to develop functions of data to the greatest extent and maximize the value of data.

As a traditional way of analyzing data, statistical analysis has been comprehensively studied and is still widely used nowadays \cite{tsou2014community, steurer2013acquaintance}. For example, Hasan et al. \cite{hasan2013understanding} analyzed urban human mobility patterns by determining and comparing a set of statistical human properties based on check-in data from online social media. The statistical features they used are the timing of visiting different places depending on category, and frequency of visiting a specific place. Vitak \cite{vitak2014facebook} analyzed the statistical features of Facebook users' interaction data to figure out how people maintain relationships with others.

However, nowadays datasets are becoming bigger, changeable, and diverse. We are entering the emerging field of big data. Traditional probability-based statistical cannot meet the requirement of processing such a large data set. This deluge of data calls for automated methods of data analysis, which is what machine learning majors in. Machine learning can be defined as a set of methods that can automatically detect rules in data, and then to predict future data or perform other kinds of decision making under uncertainty. In the paper, we follow the major classifications of machine learning, and review the algorithms used to solve the above issues derived in the field of data-driven CSS from three aspects, which are supervised learning, unsupervised learning, and semi-supervised learning, as shown in Fig.~\ref{dataanalysis}.

\textit{1) Supervised Learning}

Supervised learning needs a labeled training dataset, consisting of a set of training examples. All the examples are paired with input-output samples. That is, each pair of examples consists of an input object and the desired output value and the output values are in a known range. The supervision in the learning comes from these labeled examples. The algorithm of supervised learning analyzes the training data, producing an inferred function. This function is used to map new input examples to output class. Supervised learning can be divided into classification and regression according to the type of output class. When the output class attribute is discrete, it is the classification; on the contrary, when the class attribute is continuous, it is the regression.

\textit{a. Classification }

The purpose of classification is distinguishing objects of different classes or predicting the class label of new records. Classification technology is used for binary and nominal categories. The ordering among the values of categories has no meaning. In another word, it does not consider the order among the categories.

The target function namely classification model learned from the training data set is called classifier. There are many kinds of classifiers, such as Decision Tree learning, Bayesian classifier, K-nearest Neighbor classifier, Support Vector Machine (SVM), and Artificial Neural Networks (ANN) classifier. Based on neural network technologies, deep learning algorithms recently have been widely developed and utilized \cite{Mu2019deep}. Researchers have proposed many improved deep learning algorithms in solving issues such as image recognition, academic recommendation, and decision making. Apart from deep learning methods, the commonly used classifiers in CSS are Decision Tree learning, Bayesian classifier, and SVM. Therefore, we introduce these three kinds of classifiers here. Except for these three kinds of classifiers to the mainstream, there are also some other classifications like AdaBoost \cite{cranshaw2010bridging} which we do not introduce in detail.

\begin{itemize}
\item Decision Tree. A Decision Tree is constructed from the training data set, which is like a tree structure. In the Decision Tree, each internal node contains a test on an attribute to separate records that have different characteristics, each branch represents an outcome of the test, and each leaf node is assigned a class label. When classifying a test record, we need to apply the tree condition to the record and follow the appropriate branch until we get the leaf node as the final class.

There are many Decision Tree algorithms that have been implemented. Iterative Dichotomiser-3 (ID3) Classification and Regression Trees (CART) are the earliest algorithms designed for Decision Tree. And C4.5, the successor of ID3, became a benchmark that is often compared to newer algorithms. Chittaranjan et al. \cite{chittaranjan2011s} presented a C4.5 classifier to classify users according to their big five traits (Extraversion Agreeableness Conscientiousness Neuroticism and Neuroticism).

Based on the Decision Tree classifier, there are some forest classifiers, which consist of multiple trees. For predicting whether the friendships exist or not, Cranshaw et al. \cite{cranshaw2010bridging} implemented the Random Forests classifier which randomly constructs a forest using several Decision Trees without relevance as the basic classifier.

\item Bayesian classifier. Bayesian classification is based on Bayes' theorem. Bayesian classifiers are statistical classifiers that predict the probability that a given test belongs to a certain class. They select the class with the maximum probability as the final one which the test record belongs to.

The most typical Bayesian classifier is Naive Bayes Classifier which is also known as a simple Bayesian classifier. Naive Bayes is also applying Bayes' theorem but with strong independence assumptions between the features. That is, when the class attribute values have been given, other feature attributes and conditions become conditionally independent. For practical application, Singh et al. \cite{singh2013predicting} used a Naive Bayes method to predict the high or low propensity of couples to demonstrate a specified spending behavior (overspending, loyalty, and diversity). Furthermore, in the Naive Bayes classifier, there is an assumption of class conditional independence. But in reality, dependencies can exist between these features. Bayesian Belief Network, also known as Belief Networks or Bayesian Networks, is effective in this condition.

\item SVM. Belonging to the linear classifier, SVM searches for the linear optimal separating hyperplanes to separate data from different classes. When the training data is nonlinearly separable, SVM uses a nonlinear mapping to transform the original training data into a higher dimension. Then it constructs a hyperplane or a set of hyperplanes in the new dimensional space. So SVM can be used for both linear and nonlinear data.

SVMs are also widely used classifiers in the CSS analysis. Yu et al. \cite{yu2013understanding} adopted the SVM approach to predict friendship based on mobile phone data, recognizing reciprocal friends, non-friends, and non-reciprocal friends as well. Chen et al. \cite{chen2013friendship} proposed an SVM-based friendship prediction, assuming that a non-linear boundary exists among the friend pairs and non-friend pairs. Montjoye et al. \cite{de2013predicting} used the SVM classifier model to predict whether phone users were low, average, or high in the five personality dimensions (neuroticism, extraversion, conscientiousness, agreeableness, and openness).
\end{itemize}

\begin{table*}
\centering
\caption{Typical classifiers.}
\label{tab:data}
\begin{tabular}{|m{3cm}<{\centering}|m{5cm}<{\centering}|m{5cm}<{\centering}|}                                               \hline
Classifiers   & Applications   & Characteristics              \\ \hline
Decision Tree         & Classifying personalites and comments \cite{chittaranjan2011s}; Predicting friendships \cite{cranshaw2010bridging}       &Rule based model; Sensitive to interactions; Allowing different data types; Disadvantage for online learning                  \\ \hline
Bayesian Classifier        & Predicting propensity and link strength \cite{singh2013predicting}; Modeling motion patterns \cite{sadilek2012finding}         & Probabilistic model; Low algorithm complexity; High accuracy and efficiency                \\   \hline
SVM        & Predicting friendship patterns \cite{yu2013understanding, chen2013friendship} and personalities \cite{de2013predicting}        &Statistical method; Small number of samples; High accuracy; Low training speed                    \\ \hline
\end{tabular}
\end{table*}

We summarize these classifiers in Table~\ref{tab:data}. As mentioned in the table, these classifiers have different characteristics. Decision Tree is a rule-based model with convenience to notice the interactions among the features. The types of the data in Decision Tree can be different while other classifiers need uniform data types. However, when adding new examples to the training data, the Decision Tree needs to be updated completely, so it has a disadvantage of online learning. Bayesian classification is probabilistic with low algorithm complexity. When applied to large databases it exhibits high accuracy and speed. Nevertheless, it cannot learn the interactions of the features. SVM is a statistical method. SVM requires a relatively small number of samples for training and has high accuracy. But it needs much memory and the training speed can be extremely slow. Researchers must choose the most proper classification method in the study process by considering the various characteristics of these classifiers.

When doing the research, we can apply more than one classifier in the model to find the most suitable one. Or we can compare the effectiveness of different classifiers in the same research issues. The evaluation indexes conclude the accuracy of classification, the running speed of the algorithm, and the fitness to different data sets. Min et al. \cite{min2013mining} used a model with pairwise SVMs to evaluate how well the life facets (family, work, and social) could be classified through different smartphone data sets and compared the effectiveness with the Decision Tree(C4.5) and Naive Bayes by accuracy. The above algorithms, such as C4.5, Random Forests, Naive Bayes, SVM, and so on, have all been implemented in WEKA (Waikato Environment for Knowledge Analysis) which is a free software written in Java. And researchers mainly use it to do classification analysis. Another open-source software LIBSVM is also widely used for SVMs constructing.

\textit{b. Regression }

Regression analysis is a mathematical tool for revealing correlations between one variable and several other variables. Based on a group of experiments or observed data, regression analysis identifies dependence relationships among variables hidden by randomness. Regression analysis may change complex and undetermined correlations among variables into simple and regular correlations. More specifically, regression analysis can be used to understand how the typical value of a dependent variable changes when any one of the independent variables is varied, while other independent variables are unchangeable. In regression analysis, it is possible to characterize the variation of the dependent variable around the regression function by a probability distribution.

Regression analysis is widely used for prediction and forecasting, where its use is similar to that of machine learning. It can also be used to find out which among all independent variables are related to the dependent variable. In other words, regression analysis can be used to understand the causal relationships between independent and dependent variables. Due to its great significance, it has been widely used to mine and understand human behaviors. Regression analysis can be typically divided into two categories: linear regression and logical regression.

Linear regression requires that the model is linear in regression parameters. In linear regression, data are modeled using linear predictor function, while the unknown model parameters estimated from the data. Specifically, linear regression refers to a model in which the conditional mean of $y$ given the value of $X$. A typical work that takes advantage of linear regression is about predicting consumer behavior by Goel et al. \cite{goel2010predicting}. They investigate whether search activity is a leading indicator of consumer activity by forecasting opening weekend box-office revenue, first-month sales of video games, and weekly rank of the songs. Kosinski et al. \cite{kosinski2013private} use linear regression analysis to predict human private traits and attributes from individuals' digital records. In their model the independent variable $X$ is easily accessible digital records, for example, Facebook Likes and the dependent variable $y$ is a range of highly sensitive human attributes including, sexual orientation, political view, personality traits happiness, age, gender and etc. Other applications of linear regression are to predict box-office revenues for movies from social media and to predict human dynamics in location-based services \cite{noulas2012mining}. All these studies have shown that supervised methodology based linear regression analysis may achieve a good performance in predicting human behaviors.

Another important and widely used regression analysis is logistic regression. The goal of logistical regression analysis is the same as that of linear regression, that is, to find the best fittings and most parsimonious model to describe the relationships between a dependent variable and a set of independent variables. On the other hand, what distinguishes a logistic regression from linear regression is that the dependent variables in logistical regression are binary or dichotomous. In other words, when the outcome variable is discrete, taking on two or more possible values, the logistic regression model is the most frequently used method for analyzing this kind of data. A specific form of the logistical regression model can be described as follows: $ y = \frac{e^{\beta_{0}+\beta_{1}X}}{1 + e^{\beta_{0}+\beta_{1}X}}$
Where the $\beta_{0}$ and $\beta_{1}$ have the same meaning with that of linear regression.

There are many applications of logistic regression analysis for forecasting or predicting human behaviors \cite{dong2011modeling}, especially in online social networks \cite{steurer2013acquaintance}. Steurer et al. \cite{steurer2013predicting} employ the binomial logistic regression to predict interactions between users in online social networks. In their experiments, the logistic regression achieves a better performance in predicting online interactions compared with the methods of SVM and decision trees. In \cite{oloritun2013identifying}, Oloritun et al. proposed an improved quadratic assignment procedure based on logistic regression. Their model includes reported close friendship networks as the independent variable and network of interactions as independent variables. Through comprehensive experiments, it can effectively predict the likelihood of close friendship ties.

As we presented the specific methods which belong to the supervised learning area, however, it has its own pros and cons. Supervised learning methods have several advantages: make full use of the prior knowledge on district classification, and assure the classification in advance; control the selection of training samples to improve the accuracy of classification; avoid reclassification of a spectral cluster group in unsupervised learning. However, despite the merits supervised learning owns, it also has several shortcomings: influenced strongly by subjective factors; only recognize the defined classifications.

\textit{2) Unsupervised Learning}

\newcommand{\minitab}[2][l]{\begin{tabular}{#1}#2\end{tabular}}
\begin{table*}[htbp]
\centering
\caption{Algorithms and applications of unsupervised learning.}
\label{tab:Unsupervised Learning}
\begin{tabular}{|c|c|c|c|}
\hline
Classification & Algorithms & Methods & Applications \\ \hline \cline{1-4}
\multirow{3}*{Clustering} & Topic Model & LDA & \minitab[c]{friendship prediction \cite{parimi2011predicting}, \\human patterns discovery \cite{farrahi2010probabilistic}, \\collaboration topics differentiation \cite{tang2012cross}} \\
\cline{2-4}
& Matrix Factorization & \minitab[c]{Non-negative Matrix Factorization,\\ Singular Value Decomposition} & \minitab[c]{sentiment analysis \cite{hu2013unsupervised}, \\trust relationship prediction \cite{tang2013exploiting}} \\
\hline \cline{1-4}
\multirow{3}*{Ranking} & Random Walk & \minitab[c]{Factor-based Random Walk, \\Markov Process} & \minitab[c]{friendship prediction \cite{zhang2013learning,dong2011modeling}, \\interest targeting \cite{yang2011like}}\\
\cline{2-4}
& PageRank & \minitab[c]{TwitterRank, \\UserRank and etc.} & \minitab[c]{partner selection in scientific collaboration \cite{schall2014multi},\\influential users identification \cite{weng2010twitterrank}} \\
\cline{2-4}
& Collaborative Filtering & Collaborative Filtering & partnership prediction \cite{steurer2013acquaintance} \\
\hline
\end{tabular}
\end{table*}

In machine learning, unsupervised learning is a method of trying to find hidden and intrinsic structure in data which is unlabeled and with no target attribute. It is a much less well-defined problem since precise patterns are unknown. Furthermore, unsupervised learning is arguably more typical and analogical of human learning. And it does not require a human expert to manually label the data, which makes it more widely used than supervised learning. Labeled data is not only expensive to acquire, but also it includes relatively little information, which can certainly not be enough to reliably estimate the parameters of complex models.

There are a huge number of unsupervised learning algorithms, and more are still appearing in recent years. For the research of human dynamics in data-driven CSS, the majority of unsupervised learning algorithms employed in this problem can be considered to belong to two parts; clustering and ranking. Some typical cases shall be introduced in detail as well as their variants and extensions, whose specific algorithms and applications are shown in Table~\ref{tab:Unsupervised Learning}.

Cluster analysis or simply clustering is the process of partitioning a set of data objects into subsets, each of which is a cluster. The objects in a cluster are similar to one another, yet dissimilar to objects in other clusters. Clustering is quite useful in that it can lead to the discovery of previously unknown groups within the data. Cluster analysis itself is not one specific algorithm, but the general problem to be solved. It can be achieved by various algorithms which differ significantly in their notion of what constitutes a cluster and how to efficiently find them. Prominent clustering algorithms leveraged in CSS contain Topic Model, Latent Variable Model, and Matrix Factorization.

Topic models provide a method to analyze and organize large volumes of unlabeled text. Intuitively, they express semantic properties of words and documents in terms of probabilistic topics, which can be considered as latent structures that capture semantic correlations among words and documents in a corpus. Topic models regard each document in a corpus as a distribution over topics, and each topic as a distribution over words. The topic model is a generative model since it designates a probabilistic way in which documents are generated.

The most widely used topic model is LDA, whose topic distribution is assumed to have a Dirichlet prior. LDA models a collection of discrete data such as text corpora. For the problem of friendship prediction, LDA was employed on user profile data with the goal of producing a reduced set of features that captured user interests \cite{parimi2011predicting}. Farrahi et al. \cite{farrahi2010probabilistic} applied LDA inference on multimodal location-proximity data to discover recurring human patterns involving the time of the day, semantic location, and proximity-based interaction type. In addition, the LDA model is highly modular and can, therefore, be easily extended. In \cite{tang2012cross}, a Cross-domain Topic Learning (CTL) approach has been proposed to learn and differentiate collaboration topics from other topics. Recently, LDA has been applied to many other fields, such as word sense disambiguation, named entity recognition, tag recommendation, and community recommendation.

Latent variable models provide an approach to find disciplines from data with the help of hidden variables. In statistics, latent (hidden) variables are not observed directly but rather inferred from other observed variables. Latent variable models aim to explain manifest variables in terms of latent variables. Such models are harder to fit than models with no latent variables, but they often have fewer parameters than models that directly represent correlation in visible space. Considering, relationship strength estimation in online social networks, Ju et al. \cite{ju2017novel} formulated a link-based latent variable model with consideration of the hidden effects of user profile similarities as well as the hidden cause of interactions between users.

There are many common methods for inferring latent variables, the typical and widely used, of which is the Expectation Maximization (EM) algorithm. EM is an iterative method for finding maximum likelihood or maximum posterior estimates of parameters, where the probabilistic model depends on latent variables. EM iteration alternates between inferring the missing values, given the parameters (E steps), and then optimizing the parameters given the filled-in data (M step). Considering human mobility prediction, EM has been employed to infer the fine-grained location of individuals, even when their data is kept private, by only accessing the location of their friends \cite{sadilek2012finding}.

Matrix factorization is the decomposition of a matrix into the product of matrices, which allows us to discover the hidden patterns underlying some data. There are many different matrix factorizations, each of which, is often applied to a specific class of problems. Non-negative Matrix Factorization and Low-rank Matrix Factorization are taken as examples. Non-negative Matrix Factorization is a method in multivariate analysis and linear algebra, where a matrix is factorized into two matrices, such that all three matrices have no negative elements. Hu et al. \cite{hu2013unsupervised} built an orthogonal Nonnegative Matrix Tri-Factorization based model for unsupervised sentiment analysis with emotional signals. Low-rank Matrix Factorization is a minimization problem that produces matrices with reduced rank. In considering trust relationship prediction, Tang et al. \cite{tang2013exploiting} proposed an unsupervised framework hTrust, which incorporated low-rank matrix factorization with homophily regularization.

For prediction or recommendation \cite{wang2015friendbook, lee2017improving}, outcomes ranking, which can evaluate complex information according to certain criteria, through reducing detailed measures to ordered sequences. Some ranking-related algorithms include Random Walk \cite{Feng2019Random}, PageRank as well as Collaborative Filtering.

A random walk is a process in which randomly-moving objects wander away from the point of origin. To evaluate the power of transitivity in friendship prediction, Zhang et al. \cite{zhang2013learning} designed LFPN-RW which modeled the friend-making behavior as a random walk upon the latent friendship propagation network. The proposed friendship-interest propagation framework in \cite{yang2011like} devised a factor-based random walk model for interest targeting and friendship prediction. Various types of random walks are of wide utilization, and the common one seems to be the Markov process. It is a stochastic process that changes its state only depending on the current state, and for the discrete state and continues time, here comes the Markov jump process. Considering friendship prediction, Dong et al. \cite{dong2011modeling} described a Markov jump process to capture the co-evolution between social relationships and individual behaviors, in terms of the rates of visiting places and friends.

PageRank is a link analysis algorithm, which assigns a numerical weight to each element of a hyper-linked set of documents, with the purpose of ranking its relative importance within the set. It is now regularly applied in many different research fields, such as bibliometrics and social network analysis, link prediction, and recommendation \cite{wang2017exploiting}. Schall \cite{schall2014multi} proposed a novel topic-sensitive authority model that is based on a unified PageRank model to evaluate the importance of organizations, in order to select partners in collaborative scientific environments. To identify influential users in an online social network, the extension of PageRank algorithm TwitterRank \cite{weng2010twitterrank} was proposed, which utilized link structure and content analysis to find influential users on Twitter. Additionally, another popular ranking algorithm is Collaborative Filtering, which is a process of filtering for information or patterns with collaboration among multiple agents or data sources. Collaborative Filtering in \cite{steurer2013acquaintance} provided a method to find the feature which offered the highest information gain to predict partnership in online and location-based social networks.

Although we presented the several effective methods which belong to the unsupervised learning area above, it has its own pros and cons either. The major advantages are: only need certain knowledge to explain the classified cluster group; reduce the chance of human errors; unique categories can be identified. On the contrary, it also has several shortcomings: need a large amount of calculation on results to obtain reliable classifications; difficult to match the cluster groups to corresponding classifications.

\textit{3) Semi-supervised Learning}

Semi-supervised learning lies between unsupervised and supervised learning. It refers to learning problems in which a small amount of labeled data and a large amount of unlabeled data exist. These problems are very natural, especially in domains where the acquisition of data is cheap but labeling data is time and expertise intensive. Moreover, many researchers have found that unlabeled data, when used in conjunction with a small amount of labeled data can produce a considerable enhancement in learning accuracy. In this situation, semi-supervised learning can be of great practical value, but it has not been widely used in the field of CSS.

At present, most semi-supervised learning algorithms are based on extending unsupervised or supervised learning to include additional information, typical of the other learning paradigm. Particularly, semi-supervised learning contains several different settings, including semi-supervised classification, constrained clustering as well as regression with partially labeled data, and so on. During the past couple of years, the most active area of research in semi-supervised learning has been in graph-based methods. These graph-based methods can properly depict the structural characteristics of data itself, however, it needs a great amount of calculation and may become a barrier to practical application.

Various graph-based algorithms for semi-supervised learning have been proposed in the recent literature in the field of CSS. Graph-based semi-supervised learning starts by constructing a graph, where nodes are instances of the labeled and unlabeled data, and edges represent similarities between instances. Known labels are used to propagate information through the graph in order to label all nodes. Many researchers employed graph-based semi-supervised learning algorithms to achieve higher accuracy in terms of prediction and identification. Li et al. \cite{li2017understanding} presented the initial investigation of the time delay in reciprocal relations from Tumblr that consists of 62.8 million users and 3.1 billion users following relations with a timespan of multiple years. Zhang et al. \cite{zhang2015influenced} incorporated social influence locality into a factor graph model to study the social influence problem in a large microblogging network.

\subsection{\textbf{Validation}}
A convincing research work consists of not only the proposed new methods or mechanisms, but also needs the proof of its merits and validity. In other words, the performance of a model should be checked and proved before publications. In the validation process, the model induced from the collected data should be substantiated on whether it exactly solve the issue and how efficient it is. After the validation process, the proposed method can be determined to be effective or not. In other words, validation is the last step of the research procedure and is the guarantee of a well-performed model. The validation can be conducted from the perspective of both internal and external. The internal validation refers to how well an experiment is done, especially whether it avoids confounding. And external validity refers to how well data and theories from one setting apply to another. Here some usual benchmarks and methods of external validation are introduced as follows.

\textit{a. Evaluation of Classifiers}

In the above section, we comprehensively introduce the data analysis methods used these days, of which the main purpose is classifying the data as we called classifiers. Diverse classifiers are utilized to analyze various kinds of data sets, and the follow-up studies depend heavily on the analytical result. In other words, the performance of classifiers is crucial to the whole research work. Therefore, several classifiers' evaluating methods will be introduced in the following.
\begin{itemize}
\item Holdout Method. In holdout method, it divides the original data with labeled examples into two disjoint sets, which are training and testing sets. The training set is used to obtain the classifying model, and its performance is evaluated on the testing set. The accuracy of the classifier can be estimated based on the induced model's accuracy, which is according to the test set. Besides the differences in the proportion of reserved data for training and testing set, researchers in \cite{yu2013understanding, tang2012cross} applied this method for validation.

\item Random Subsampling. Multiplicating the holdout methods to improve the estimation of a classifier's performance is the random subsampling method.

\item Cross Validation. It first partitions the original data into two subsets with equal size, then chooses one set to be the training set and the other one to be testing set to proceed the holdout method. Subsequently, switch the role of the two subsets and estimate the accuracy again. Cross validation is one of the most commonly used validation methods, for instance, it is applied in \cite{choi2013mining, bischoff2012we} to verify their analytical results.

\item Bootstrap. The above methods are all assuming that there is no replacement on the selection of training set, while for a bootstrap method, it exits replacement.
\end{itemize}

\textit{b. Comparison of Classifiers}

There always exist other classifiers targeting the same issues and are proved to be convincing with high accuracy. So comparing the results of current methods with existing comparable and validated models (if any) is an effective method available. On the other hand, if two new proposed models of the same puzzle lead to different or even opposite conclusions, then an examination is needed. Another benefit of comparison is that different models based on different paradigms can provide a more comprehensive perspective for the researchers. Generally, we can use three statistic methods to compare the performance of different classifiers, which are estimating a confidence interval for accuracy, comparing the performance between models, as shown in \cite{chen2013friendship}.

\textit{c. Predictive Validation}

If the model is used to do prediction, in other words, the purpose of the model is to predict future behavior or future event, and then comparisons can be made between the tracking data and the model¡¯s forecast. After that, the determination will be made about whether the model is effective or not. The values of accuracy are also computed during this process which can be used as a standard to judge a model whether good or not.

\textit{d. Participatory Approaches}

Participants who produced the data or stakeholders such as model users are best qualified to speak on the effectiveness and veracity of the model. Online surveys or interviews toward related individuals can be made to investigate whether the model is satisfactory. But this approach performs well only when the scale of related participants is not large, as shown in \cite{ortigosa2014sentiment}.

\section{Open Challenges and Future Trends}

In previous sections of this paper, the state-of-the-art research progresses involving human dynamics in the field of CSS are reviewed. Individuals, relationships, and collectives are delved into the survey which is not projected to be exhaustive, but we try to cover the majority of categories of related research efforts. Besides the research foci demonstrated above, there are still many open issues which are representatives of interesting research directions both at the theoretical and the applied levels. We give a non-exhaustive, subjective list of such problems that seem particularly potential for further study in this section.

\subsection{Interactions Between Online and Offline}

Recent research mostly focuses on how offline behaviors affect users' online behaviors. But with the wider use of the Internet, people are spending increasing time on the virtual world to get more information and interact with each other from online lives. Even in some cases, people make more direct contact with electronic equipment than communicating with others in reality. Therefore, the influence of online activities on offline lives needs to be deeply mined. Besides, there exist many investigations of identifying users' states including physical states and mental states by users' online behaviors or some other data getting from users' mobile devices. However, previous research ignores the dynamic nature of human states and the dynamic and changing processes are even more significant to understand human behaviors.

\subsection{Data Quality and Analysis Validation}

The primary data-related issue is how to encourage people to share their data. The data used in CSS research is highly associated with people's private lives. People may be confronted with privacy issues when sharing personal data. Therefore, privacy protection is needed in the research process. Data anonymity is one possible solution for such issue, which can avoid the leakage of private information. Another method is designing various experiments and letting participants themselves decide whether their information can be used or not. A systematic method guaranteeing data availability and data privacy is the fundamental base of the whole research process.

Another issue is how to provide high-quality data, which contains three aspects. The first one is avoiding mistakes and ensuring accurate statistics in data collection. The second one is extracting high-level intelligence data from large-scale useless sensing data. And the last one is managing heterogeneous data sources and combining data of different aspects effectively. These are also important emphases of CSS.

In addition, one important issue we need to consider is the analysis validation \cite{lazer2014parable}. In February 2013, the prediction result of Google Flu Trends based on the estimation of surveillance reports from laboratories across the United States, was the double of the result of the Centers for Disease Control and Prevention. This error alerts us with two significant aspects we need to consider when solving problems in the future. The first one is whether the data we plan to use conforms to our designed experiments or instruments since using estimated data source can lead to unexpected results. Another factor is the dynamics of algorithms, whose stability and comparability may have crucial influences on the final outcomes.

\subsection{Relationship Evolution}

New technologies offer a minute-by-minute description of interactions over expanding periods of time and provides information on ever-changing relationships. As a result, relationship evolution attracts more and more researchers' attention. Relationship attributes, including types and strength of relationships, are evolving as time progresses. Various kinds of relationships exist in our lives such as marriage, neighborhood, work, information transfer, exchange, co-membership, and so on. Some relationships can convert into other relationships. For example, co-membership is likely to become a romantic relationship over a period of time. Moreover, the strength of relationships is varying (e.g., attenuation or enhancement) through online communications and offline interactions.

Additionally, some social theories of relationship deduced from traditional investigations and statistics have not been sufficiently verified and proved due to technical restrictions. However, the vast and longitudinal human interaction datasets not only offer qualitatively new perspectives on relationship evolution, but also provide reliable evidence for the existing social theories. Accordingly, it is a promising direction to take advantage of real-time interaction data to study how relationships evolve and how social theories work under the dynamics of relationships.

\subsection{Social Dynamics of Science}

Academia is a special community that is composed of scholars. With the development of science, scholars are producing more and more data including millions of publications, citations, figures, and large-scale related data such as academic social networks, slides, course materials. One interesting issue is how to analyze these data to better understand the rule of science itself. By mining scholarly data, we can solve critical problems in academia and provide personalized services for scholars. Understanding how research topics emerge, evolve, or disappear, what are the trends in a field, who are the experts in a given area, and how to recommend papers, journals, and collaborators to researchers are some of the major topics of the rapidly emerging field involving scholarly big data. At the same time, an important question under the dynamics of science is how to utilize recent technologies such as data mining and machine learning to gather and analyze large volumes of scholarly data to better study science itself and scholars ourselves.

\section{Conclusion}
CSS has been a topic of research interest in recent years due to an increasing realization of the enormous potentials of its data-driven capabilities. The availability of unprecedented amounts of data on human interactions in different social spheres opens the possibility of transcending existing social behavior boundaries. These data can be used to validate the results of simulation models and socio-economic theories. Some relevant conclusions may be applicable to a greater proportion of individuals. However, if included in the modeling design, it may yield profoundly new insights.

In view of this, it is apparent that the analysis of the data on the daily activities of individuals will certainly contribute to the understanding of human dynamics mechanisms. In this paper, we first summarize typical problems in human dynamics from the view of individuals, relationships, and collectives. Then, the most commonly used methods which may benefit the study of the above issues are presented. Furthermore, several promising research areas and open issues are discussed.

\appendices

%\section{}

%\section*{Acknowledgment}


\ifCLASSOPTIONcaptionsoff
  \newpage
\fi



\bibliographystyle{IEEEtran}
\bibliography{IEEEabrv}
\end{document}